%% file: paper_v2.tex
\documentclass[sigconf,nonacm]{acmart}

\usepackage{algorithm}
\usepackage{algorithmic}
\usepackage{newfloat}
\usepackage{listings}

\floatstyle{ruled}
\newfloat{listing}{tb}{lst}{}
\floatname{listing}{Listing}

\copyrightyear{2027}
\acmYear{2027}
\setcopyright{acmlicensed}
\acmConference[WWW '27]{Proceedings of the ACM Web Conference 2027}{2027}{TBD}
\acmBooktitle{Proceedings of the ACM Web Conference 2027 (WWW '27)}
\acmDOI{10.1145/nnnnnnn.nnnnnnn}
\acmISBN{978-x-xxxx-xxxx-x/YY/MM}

\input{preamble}

\newcommand{\ownershipSourceWikidata}{\Res{legacy.wikidata}}

\title[Measuring Media Concentration Across \Res{frame.ranked} Countries and Seven AI Systems]{Who Owns Digital News, and Who Gets Cited: Measuring Media Concentration Across \Res{frame.ranked} Countries and Seven AI Systems}

\author{Peter Carragher}
\affiliation{%
  \institution{Carnegie Mellon University}
  \city{Pittsburgh}
  \state{Pennsylvania}
  \country{USA}}
\email{pcarragh@andrew.cmu.edu}

\author{Kathleen M. Carley}
\affiliation{%
  \institution{Carnegie Mellon University}
  \city{Pittsburgh}
  \state{Pennsylvania}
  \country{USA}}
\email{carley@andrew.cmu.edu}

\begin{document}

\begin{abstract}
Digital news outlets compete for audience attention. Search engines equipped with generative AI systems such as Google's AI overviews increasingly decide where that attention goes, as well as the revenue it represents.
Yet, the lack of a consistent data collection infrastructure between national-level regulatory bodies has hindered between-country comparative analysis of media concentration, and with it any assessment of how these intermediaries reshape digital news markets.
To address this issue, we compile an ownership dataset for \Res{frame.withOwner} digital news outlets drawn from Wikipedia and Wikidata, achieving high agreement with independently compiled US and UK news directories.
Using this dataset, we derive traffic-weighted concentration indices for \Res{frame.ranked} countries.
We find that journalists report more owner pressure in countries with higher concentration ($\rho = $\Res{wjs.hhi.rho}), and that outlet-level measures understate concentration, with owner-level indices a median \Res{frag.ratioMedian}× higher.
Applying concentration measures to AI-mediated news discovery, we find that citation distributions from major AI systems are up to \Res{ai.ratioMax}$\times$ more concentrated than traffic,
and that if Google AI Overviews mediated all organic search, the resulting rise in US concentration would be several times that associated with highly regulated media mergers and acquisitions.
Finally, our analyses indicate that AI systems preferentially cite state-controlled media outlets when prompted in that country's native language.
Together, these findings raise concerns about the effect AI systems may have on digital media.
We release the dataset to enable the monitoring of digital news globally, so that the impact of disruptive technologies on digital media markets can be tracked as they emerge.
\end{abstract}



\maketitle

\section{Introduction}

On October 6, 2026, despite being challenged by twelve state attorneys general and the Writers Guild of America, Paramount Skydance acquired Warner Bros.\ Discovery in a deal worth over \$110bn.
By placing CBS News and CNN under a single owner, critics are concerned the merger will impact external pluralism in US news.

When questions of media concentration invariably arise in such cases, they are answered anew each time, a process made difficult by a lack of standardization in how concentration is measured.
National regulators count licenses, circulation, or revenue within their own borders and on their own definitions.
As a result, there is no readily available dataset that can be used to compare media concentration between countries.
This is a particularly timely issue, given that the technological changes actively shaping the news media, such as algorithmic curation and the proliferation of AI overviews in search engines, are global in nature.

\begin{figure}[t]
\centering
\includegraphics[width=\linewidth,clip,trim={0 0.8cm 0 0}]{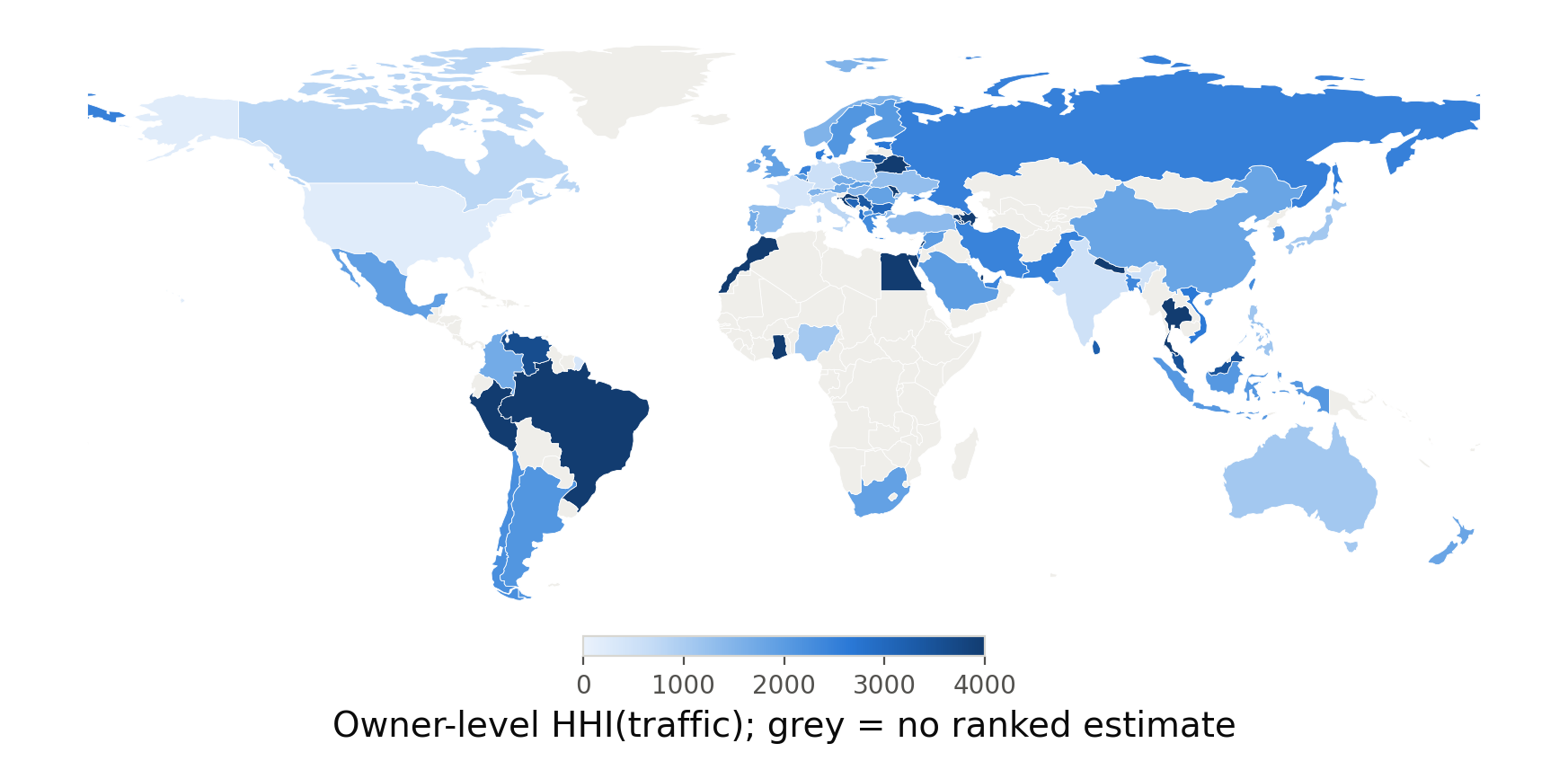}
\caption{Ownership concentration by country, as HHI on traffic shares by owner. Higher values mean traffic is held by fewer owners.
Unshaded countries have no estimate. Each country's state-controlled media are grouped under one state entity (state-sponsored are not).}
\label{fig:worldmap}
\end{figure}

Digital news outlets exist in a market. Outlets compete for audience attention online,
converting referral traffic from search engines and, increasingly,
generative AI systems into revenue. Missing ownership data therefore leave a Web
economics question open: how concentrated are national markets for digital
news traffic once websites are grouped by owner, and are citations from AI
systems more or less concentrated than that traffic? 
This places our work, along with its recommendations regarding competition policy for digital platforms, within the Web Economics and Digital Society track.


To answer it, we construct an international news ownership dataset from crowd-sourced data on Wikipedia. We assign ownership labels to \Res{frame.withOwner} domains and compute traffic-weighted concentration
indices (Herfindahl Hirschmann (HHI)) for
\Res{frame.ranked} countries, while the largest prior cross-national study covered 30 \citep{noam2016media}.

The picture that emerges is one of substantial variation (\figref{fig:worldmap}). Here, HHI is the sum of squared market shares and runs from near 0 when many small owners split the market to 10,000 for a single owner. Its reciprocal ($10{,}000/\mathrm{HHI}$) is the effective number of equally sized owners, and competition regulators use it to assess market concentration and the effect of mergers \citep{hill2006measuring}. In our measurement, the number of effective owners per country ranges from more than ten to fewer than three. 
Despite this variation, grouping outlets by owners reveals higher traffic concentration levels, with owner-level indices being a median factor of \Res{frag.ratioMedian}$\times$ outlet-level indices.

We validate our concentration indices by comparing against previously published concentration figures \citep{noam2016media}, and both an independently compiled US ownership directory \citep{LocalNews2025} and a UK news directory \citep{public_interest_news_foundation}. We further compute agreement between the dataset and a set of hand-annotated is-news labels, and between each outlet's assigned country and where its audience is located \citep{similarweb}. Our indices reveal that media concentration tracks conditions reported \citep{hanitzsch2019worlds} from inside newsrooms:
journalists report more owner and commercial pressure where ownership is concentrated.

That excessive media concentration can result in such harm is the reason that the FCC regulates traditional mergers and acquisitions.
However, existing regulations have blindspots around emerging technologies, and AI-mediated news discovery is already redistributing news attention. AI-generated answers have been shown to substantially reduce click-through to publishers \citep{agarwal2026impact}, and publishers are rewriting distribution and search strategy as they compete to gain AI citations \citep{ponsford2026journalists}.
Collectively called Generative Engine Optimization (GEO) after Search Engine Optimization (SEO), these emerging strategies attempt to increase the rate at which a domain is cited by generative models \citep{aggarwal2024geo}, such as those deployed in AI overviews \citep{chen2025generative}.

Whether the distribution of citations from AI systems follows existing media market structures, and what impact those systems will have on future concentration levels, are open questions.
We apply our ownership dataset to investigate this, using citation counts
from seven AI systems: in doing so, we find that AI citation distributions are
\Res{ai.ratioMin}$\times$--\Res{ai.ratioMax}$\times$ more concentrated than
traffic distributions at the outlet-level, and \Res{ai.ownerRatioMin}$\times$--\Res{ai.ownerRatioMax}$\times$ more concentrated at the owner-level.
In addition, we find that the AI amplification effect (i.e. the relative increase in concentration in citations over traffic) is most pronounced in countries with lower traffic concentrations.
As such, if traffic to news outlets shifts to follow citations from AI systems, the traffic concentrations reported here will increase, with less concentrated markets being disproportionately affected.
However, given that measuring the impact of AI citations on traffic directly requires longitudinal data, we leave this to future work.

Underscoring the importance of these findings are a series of copyright infringement and unfair competition law suits that news media companies are beginning to bring against AI intermediaries over the impact of AI systems on declining news outlet traffic and advertising revenues \citep{tobitt2026news}.
Whether regulators should focus their efforts on these new cases or on those associated with traditional mergers and acquisitions becomes an issue of policy priority. Does a merger of two of the largest broadcast news owners move a country's media concentration more, or less, than the introduction of AI overviews in Google Search? We estimate the impact of both against our indices and return to the comparison in the Discussion.

\section{Related Work}

\paragraph{Digital News}
Research on digital news covers a broad range of topics such as local news provision \citep{bisiani2025uktwitnewscor}, misinformation \citep{carragher2024detection} and ``pink slime'' detection \citep{aljebreen2024analysis}, revenue modeling for unreliable news \citep{papadogiannakis2023who}, cross-national news diversity measurement \citep{chen2024global}, and social media based news exposure \citep{chouaki2024what}. While these works classify and track digital news outlets in various ways, they do not record who owns them. The resulting lack of market-structure analysis for digital news media has been identified as a research gap in a recent literature review on organizational influences in newsrooms  \citep{theine2025does}.

\paragraph{Media concentration measurement}
Increases in U.S. media concentration between 1983, when 50\% of the US media was owned by some fifty companies, and 2000, by which time this had fallen to six companies, motivated a large body of research \cite{bagdikian2004new}.
Researchers have developed several measures to quantify concentration, such as the Noam \citep{noam_media_2009} and Herfindahl Hirschmann (HHI) indices \citep{hill2006measuring}.
These measures have enabled the effects of mergers and acquisitions to be evaluated, such as by the FCC who adopted HHI in 2005 \citep{hill2006measuring}.
Since then, measures of media concentration have been applied to media environments around the world in which corporate consolidation is problematic \cite{noam2016media,lidberg_distortion_2019,sjovaag_media_2019,trappel_soaring_2022}.

\paragraph{Comparative media analysis}
Comparative media analysis describes how various journalistic cultures emerge given differing political systems \cite{hallin2004comparing}.
We draw on three cross-national survey-based instruments: the Worlds of Journalism survey (WJS) \citep{wjs3reports}, the Reuters Digital News Reports' country-level media trust scores \citep{egan2026reuters}, and Reporters Sans Frontières (RSF) measures of press freedom \citep{rsf2026}.
Based on these indices, it has recently been demonstrated that LLMs exhibit pro-government valences when prompted in the languages of countries with lower media freedom \citep{waight2026state}, a finding we build on in this work.

\paragraph{Search, AI intermediaries and publisher economics}
Search engines and AI overviews act as intermediaries between publishers and their audiences. AI-generated answers reduce click-through to publishers \citep{agarwal2026impact}, generative engines can be optimized for citation \citep{aggarwal2024geo,chen2025generative}, and publishers have begun to litigate over lost traffic and advertising revenue \citep{tobitt2026news}. Search engine audits have frequently measured which sources results pages and news search surface \citep{robertson2018auditing,le2019measuring,urman2021auditing} and how much traffic search sends to state-linked sites \citep{spangher2020characterizing}. Recent work extends such measurement to LLM-based search engines whose source selection can be manipulated \citep{chen2026unveiling} when retrieving citations for new user queries \citep{sayyidali2026are}.

A parallel line of research develops diversity-aware news aggregation and recommendation algorithms \cite{chakraborty_editorial_2018,abdollahpouri_toward_2021,choi_not_2022} by measuring external pluralism and incentivizing it through algorithmic intervention \citep{vrijenhoek2021recommenders,vrijenhoek2022radio,scalvini2023making}.
Although we do take advantage of datasets of AI citations to news outlets \citep{miroyan2026search,yang2025news}, to the best of our knowledge, there are no works that investigate the impact of AI systems on media concentration. \citet{pierson2023governing} emphasize that current AI media policy research has largely ignored structural power issues, and explicitly list the impact that AI has on trends of “increasing market concentration and power asymmetries” as a blindspot in the literature.

\section{Dataset}
\label{sec:dataset}

To compile our dataset of global news sources and their owning companies, we utilize crowdsourced lists of broadcasting company assets on Wikipedia. We compiled a list of broadcasting companies for which a Wikipedia category exists\footnote{e.g., Nexstar's page: \url{https://en.wikipedia.org/wiki/Category:Nexstar_Media_Group}} and used Wikipedia's Petscan service to retrieve the Wiki-pages of the news outlets each company owns. News sources can be newspapers, TV stations, and radio stations.
To restrict to outlets with an official website, we query Wikidata via SPARQL for each outlet's official domain (\ownershipSourceWikidata{} outlets) and fall back to the website link in the outlet's Wikipedia Infobox. This yields \Res{frame.domains} news domains across \Res{frame.countries} countries.

We combine these ownership records with SimilarWeb traffic estimates and site
categories. Previous research has validated SimilarWeb traffic estimates against estimates from both Ahrefs and computations of PageRank over the CommonCrawl webgraph \citep{carragher2025misinformation}. While the definition of what counts as news is subject to extensive research \citep{harcup2017news}, we simply use SimilarWeb site categories to filter out non-news websites from the dataset.
Specifically, to the list of \Res{frame.domains} news domains, we remove domains that: are missing a country code, have a traffic estimate of zero, have a SimilarWeb site category unrelated to news (removes platforms, commerce and streaming sites),
or are news aggregators (via a Wikidata entity-type rule). Of the remaining
\Res{frame.withTraffic} domains, we filter out any that do not have an ownership record. The resulting dataset has \Res{frame.withOwner} domains.
We annotated a sample of $n=200$ of the remaining domains. 75\% are news, accounting for 82.6\% of traffic in the sample. Counting magazines and sports outlets increases this to 89\% news, accounting for 95\% of traffic. The appendix details coverage, filtering, and annotation.

\begin{figure}
    \centering
    \includegraphics[width=\linewidth]{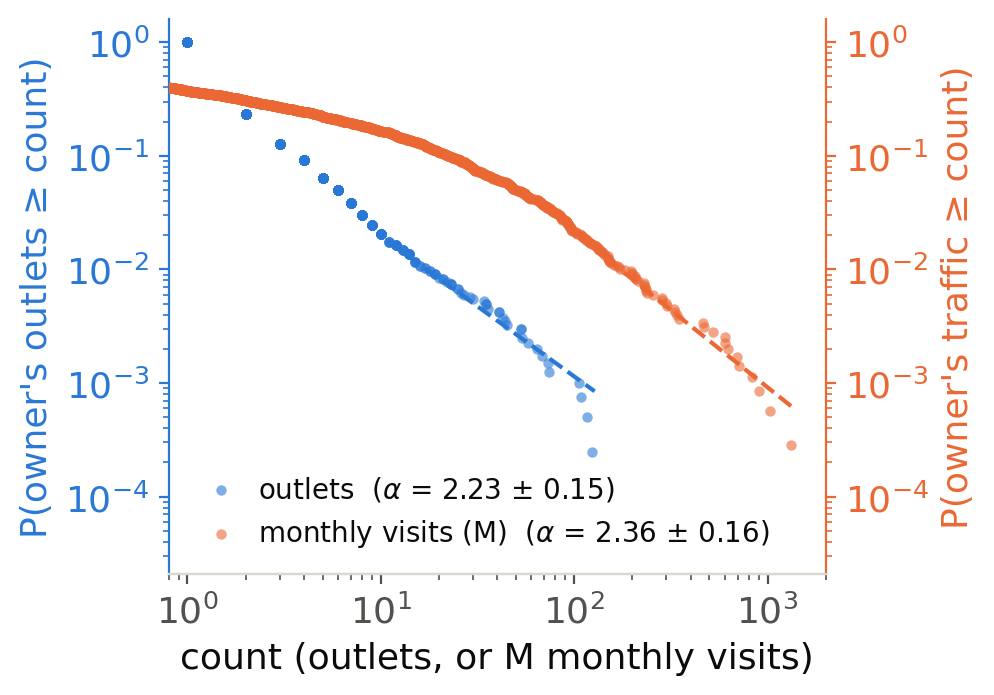}
    \caption{The ownership dataset has a long tail of owners that own many news outlets (blue, left axis) and account for large proportions of total traffic (orange, right axis; monthly visits in millions). Both follow power laws (dashed lines; $\alpha \approx 2.2$ for outlets, $\alpha \approx 2.4$ for traffic). The axis starts at 1, which hides the 63\% of owners with fewer than 1M monthly visits; probabilities are computed over all owners.}
    \label{fig:owner_size_distribution}
\end{figure}

Owner sizes follow power laws, with a small
number of large groups sitting on a long tail of smaller news organizations (Figure \ref{fig:owner_size_distribution}). Of the \Res{tail.owners} owners in the
dataset, \Res{tail.onePct} hold exactly one outlet and \Res{tail.oneOrTwoPct}
hold one or two. Together the single-outlet owners account for
\Res{tail.oneTrafficPct} of labeled news traffic, and within the median ranked
country they are \Res{tail.medianOwnerPct} of the owners and
\Res{tail.medianTrafficPct} of the traffic.

\subsection{Concentration Estimates}
\label{sec:descriptive}

For a given country with $n$ media companies, let $s_i$ represent the market share, in percent, of the $i^{th}$ largest company. We compute the largest-owner share ($s_1$), CR4 ($\sum_{i=1}^{4} s_i$), the Herfindahl--Hirschman Index ($\sum_{i=1}^{n} s_i^2$), and the Noam index (HHI$/\sqrt{n}$). A domain with $k$ co-owners contributes traffic$/k$ to each co-owner. While estimates are computed for \Res{frame.countries} countries, we report results for the set of \Res{frame.ranked} countries that have at least five identified owners. For countries that have less than five owners identified in Wikidata, coverage was insufficient to produce reliable estimates.
We compare concentration estimates for two variants of the dataset, the \Res{frame.ranked} country dataset described above (Figure~\ref{fig:worldmap}), and a smaller \Res{frame.reliable} country variant that filters out any country with less than \Res{desc.reliableMinDomains} owner-labeled news domains or whose domains account for less than
\Res{desc.reliableMinCoverage} of the traffic going to that country's news domains.

Across the \Res{frame.reliable} reliable countries the median HHI is
\Res{desc.medianHhi}, or about \Res{desc.medianEffOwners} effective owners,
with a median largest-owner share of \Res{desc.medianTopOne} and a median
top-four share of \Res{desc.medianCrFour}. The median well-covered national news
market therefore resolves to roughly six effective owners: one organization holds
about a quarter of news attention and the largest four hold $>$50\%. Across
the wider set of \Res{frame.ranked} ranked countries the median falls to
\Res{desc.rankedEffOwners} effective owners and CR4 rises to
\Res{desc.rankedCrFour}, because countries that fall outside the reliable set
are smaller and thinner. The least concentrated markets resolve to tens of effective owners, the most concentrated to fewer than three.

Concentration indices provide information beyond simply the number of
owners. The UK has \Res{desc.gbOwners} identified owners and an HHI
of \Res{desc.gbHhi}, because the BBC alone holds \Res{desc.gbTopOne} of labeled
news traffic; T\"urkiye, by contrast, has \Res{desc.trOwners} owners and a \emph{lower} HHI of
\Res{desc.trHhi}. The BBC is just one example of a public service broadcasters that accounts for significant traffic in a highly concentrated market. Among the \Res{top.countries} countries for which we identify a largest owner, that owner is a public or state entity in \Res{top.public} of them.
Outside these four, public broadcasters often act as a counterweight to an otherwise highly concentrated commercial sector.

\subsection{State-Sponsored Media Labels}
We use the state-media classification from the State Media Monitor (SMM), a hand-compiled census of state-administered news media maintained by the Media and Journalism Research Center \citep{dragomir2022state}. SMM grades outlets on ownership and governance, source of funding, and editorial autonomy.
In total, of the domains in our dataset, 298 are classified as state-controlled by SMM labels, with coverage spanning 80 countries. In our analyses, we treat all state-controlled news outlets as being owned by a single entity (i.e. the state), as treating them as independent entities would artificially inflate market competition metrics.
Wikidata supports a similar classification with high precision but low recall against SMM, so we rely on SMM and describe the Wikidata derived labels for state sponsorship in the appendix.

\section{Dataset Validation}
\label{sec:validation}

We validate the dataset by evaluating agreement with published concentration figures, independent ownership directories, and human annotations. The appendix reports further validation, such as a high level of agreement between our country labels and the country from which the majority of the news domain's audience originates, as derived from SimilarWeb data, and shows that our analyses are invariant under different inclusion criteria.

\paragraph{Published figures}
Against Noam's \emph{Who Owns the World's Media?}
\citep{noam2016media}, our index reproduces his four-firm newspaper concentration
ratio at $\rho = \Res{noam.cfour.rho}$ ($p = \Res{noam.cfour.p}$,
$n = \Res{noam.cfour.n}$). His newspaper-press HHI ranking is also reproduced
($\rho = \Res{noam.newspaper.rho}$,
$p = \Res{noam.newspaper.p}$, $n = \Res{noam.newspaper.n}$).
Noam's figures are for 30 countries in 2008--2013 and mostly measure the print press (online news is covered for only 20 countries), so moderate correlations are expected. The comparison spans more than a decade and a change of medium from newspapers to digital news. As such, our index reads systematically less concentrated than Noam's print series, which is likely related to the inclusion of a long tail of small news outlets. The rank agreement is nonetheless evidence that it tracks national ownership concentration.

\paragraph{Owner directory}
A total of $n=638$ news domains in our dataset overlap with the Medill Local News Initiative directory of US
newspapers \citep{LocalNews2025,LocalNews2024}. Against the Medill dataset, our ownership labels achieve a 73.5\% agreement rate. When weighted by traffic, the domains with matching owners account for 91.9\% of the traffic that the overlapping domains receive, revealing that agreement rates are higher for news domains that receive more traffic.

Where Medill disagrees with our dataset, we validate by manually annotating the disagreeing domains. We find that Medill is typically correct, with errors on Wikipedia being caused by the crowdsourced ownership attributes being out of date. Infoboxes often lag recent acquisitions, such as CherryRoad Media's acquisition of several Gannett newspapers \citep{GulbanCherryRoad2025}.

Medill data also contains historical data on acquisitions. Using this, we find that our Wikipedia ownership dataset reaches an agreement rate of 93.6\% on overlapping domains with the Medill dataset, if we consider either the current or last owner --- thus accounting for the Wikipedia data being out of date by a single recent acquisition. Domains matched under this condition account for 98.8\% of overlapping traffic. While this level of agreement is already very high, any error that might be induced by inaccuracies is minimized by the fact that acquisitions typically serve to increase media concentration. As such, country-level concentration measures derived from these labels would be biased in the same direction, and could be considered lower-bounds.

\paragraph{Annotations}
On \Res{annotation.n} hand-annotated domains the primary filter
has precision \Res{filter.precision}, recall \Res{filter.recall} and
traffic-weighted precision \Res{filter.trafficPrecision}, retaining
\Res{filter.retained} of the list. The unfiltered list is only \Res{annotation.newsPrecision} news by human judgement, and the non-news it contains (platforms, news aggregators, etc.) dominate traffic distributions. As such, removal of non-news websites can have a large impact on HHI measures. For example, US HHI falls by a factor of four once large platforms are removed. Finally, we annotate a subset of US news domains ($n=321$) with ownership labels and find that the Wikidata ownership labels agree with our annotations (95.8\%).

\section{Results}

We first relate our concentration indices to reports of ownership pressure \citep{wjs3reports}, then compare traffic with AI citation data from Ahrefs \citep{ahrefs} and AI Search Arena \citep{miroyan2026search}, and finally ask whether AI systems show preference toward state media \citep{dragomir2022state} and how amplification relates to media trust \citep{egan2026reuters}.

\subsection{Reported Ownership Pressure}
\label{sec:wjs}

\begin{figure}[t]
\centering
\includegraphics[width=0.9\linewidth]{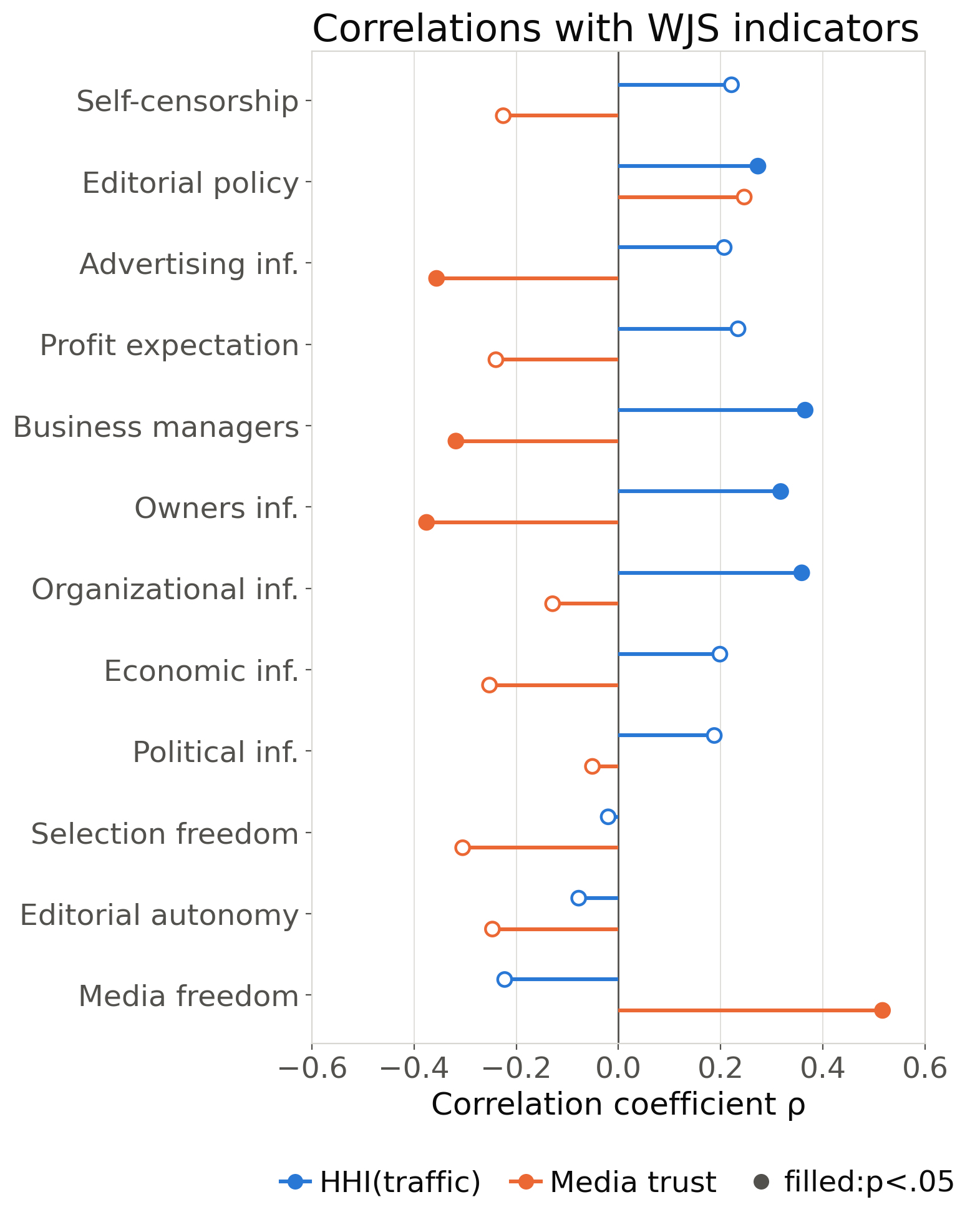}
\caption{WJS indicators show that ownership concentration (HHI(traffic)) correlates with reports of owner and manager influences by journalists \citep{wjs3reports}. The same indicators often have opposing correlations with media trust \citep{egan2026reuters}. Filled markers show significance.}
\label{fig:wjs_indicators}
\end{figure}

Table \ref{tab:wjs-pressure-vs-concentration} shows that WJS ownership pressure ratings correlate with concentration indices, in particular the largest-owner share ($\Res{wjs.top1.rho}$).
Since WJS country aggregates are produced by surveying journalists,
correlations between the WJS aggregates and traffic-weighted concentration indices (Figure \ref{fig:wjs_indicators}) represent a relation between the media market structure and the experiences of journalists in the newsroom.
We also find that this relationship is not explained by developmental indicators.
Adding controls for GDP per capita, internet penetration, and population to a regression model predicting WJS ownership pressure leaves the association intact ($\beta = \Res{wjs.olsBeta}$ on log HHI, $p = \Res{wjs.olsP}$, $R^2 = \Res{wjs.olsRtwo}$, $n = \Res{wjs.olsN}$).

\input{tables/wjs_pressure_vs_concentration}

\subsection{AI Citations}
\label{sec:ai}

This section asks whether the citations from AI systems follow the existing structure
of media ownership.
We collect Ahrefs AI-response counts showing how often each domain is cited by
one of five AI systems, ChatGPT, Copilot, Gemini, Google AI Overviews (`Google') and Grok.
Ahrefs counts AI citations based on the responses each system gives to queries drawn from Google’s “People Also Ask” corpus \footnote{\url{https://ahrefs.com/blog/brand-radar-methodology/}}.
The AI citation data covers \Res{ai.domains} news domains
across \Res{ai.panelCountries} countries.
We filter out one country for which there are fewer than 10 domains receiving AI citations, and report results on the overlap with the \Res{frame.ranked} countries for which we have reliable concentration measures. Where mean and median measures are presented over multiple AI systems, the measure is first computed separately per system, and then aggregated. Otherwise, summing two citation distributions gives an HHI that underestimates the average of their separate HHIs. \footnote{This fact follows from applying Jensen's inequality to HHI.}

Related to our research questions on the impact of AI systems on media concentration, we test four hypotheses:
\begin{description}
\item[H1] AI citations are more concentrated than traffic.
\item[H2] AI citations are more concentrated than traffic after grouping outlets by owner.
\item[H3] AI concentration tracks ownership concentration.
\item[H4] When prompted in the relevant native language, AI systems cite state media sources more than expected based on traffic levels.
\end{description}

Because the Ahrefs collection method may carry its own biases, we also compute citation counts for our news domains in the AI Search Arena dataset \citep{miroyan2026search}. Its coverage of our dataset is lower, so only Ahrefs data is used for H1--H3, but Arena adds a second measurement of Gemini, search-enabled Perplexity and GPT 4.0, and non-English prompts.

\input{tables/ai_amplification}

\paragraph{H1} \emph{Citations are more concentrated than traffic}.
For each country and each AI system, we compare the HHI of citations with the HHI of traffic over the same outlets.
A positive difference (amplification) means AI answers draw on a narrower set of outlets than audiences do.
Using a Wilcoxon signed-rank test, all \Res{ai.models} systems reject the null at $p < 10^{-6}$, supporting H1, that $\text{HHI}(\text{AI}) > \text{HHI}(\text{traffic})$.
Table \ref{tab:ai-amplification} shows that AI systems cite a narrower set of outlets than audiences visit in \Res{ai.ampPctMin}--\Res{ai.ampPctMax} of countries.

We also check whether AI citations are allocated across outlets differently from traffic.
We correlate outlets' traffic ranks with their citation ranks, and check whether the most-cited outlets are also the most-visited.
While median within-country rank correlation between traffic share and citation share indicates agreement ($\rho = \Res{ai.rankCorr}$),
this does not necessarily mean that AI citations and traffic favor the same set of outlets.

We test whether the citation ranking of two AI systems agree with each other more than their agreement with traffic estimates would predict.
The median rank correlation between each individual system and traffic is 0.51, with a mean correlation of 0.49.
If each system cited outlets independently, their correlation with traffic would account for the entire correlation between the two AI system rankings
leading to an expected correlation of $0.51 \times 0.51 \approx 0.26$.
However, the median rank correlation between two systems is 0.65.
The most-cited outlet is the most-visited in only \Res{ai.sameLeader} countries.
So, while there is substantial overlap between the outlets cited by AI and those that attract the most traffic, there are also systematic differences as exemplified by higher inter AI system agreement.

\begin{figure}
    \centering
    \includegraphics[width=0.9\linewidth]{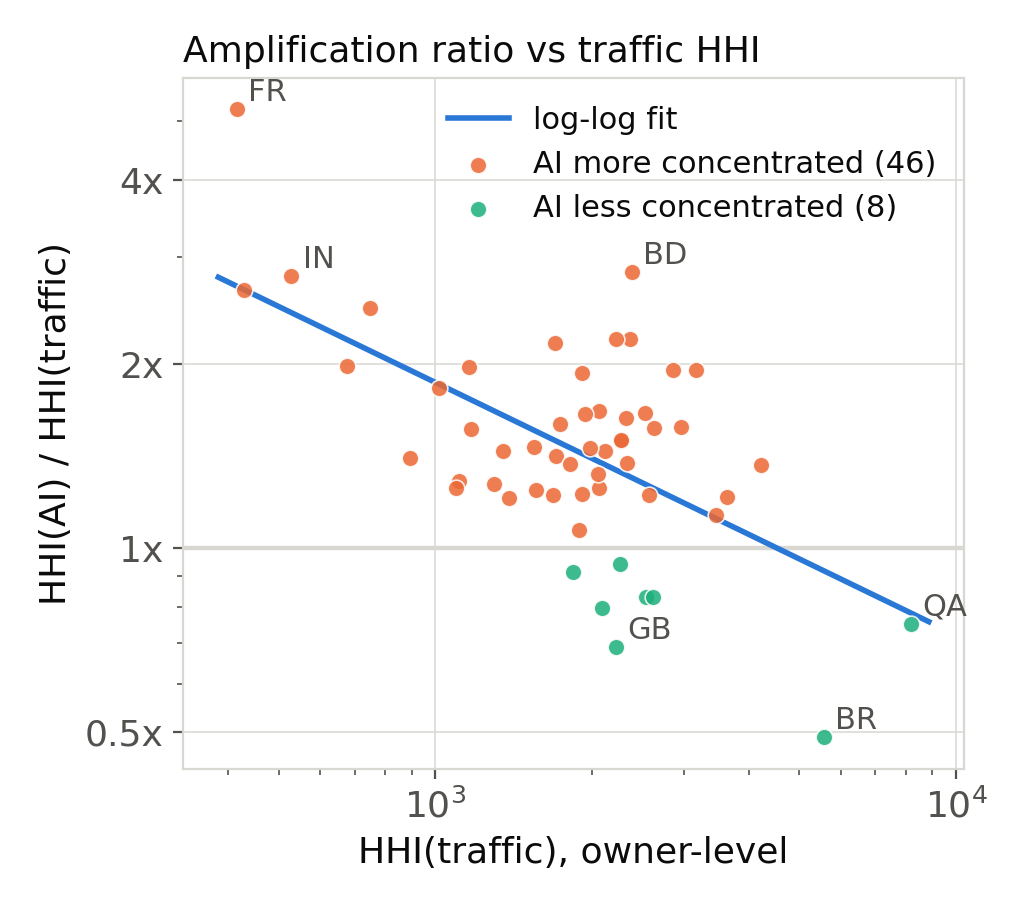}
    \caption{AI owner amplification rates are higher in countries with lower media concentration ($\rho=-0.32$*).}
    \label{fig:ai_owner_amplification_b}
\end{figure}

\paragraph{H2} \emph{H1 holds after grouping outlets by owner.}
Figure \ref{fig:ai_owner_amplification_b} demonstrates AI amplification at the owner-level, with most countries having an amplification rate $>1$.
After summing traffic and citations across each owner's outlets, we repeat the same test as for H1.
Owner-level HHI(AI) and HHI(traffic) are strongly correlated across countries ($\rho=0.78$***).
Averaged over the \Res{ai.models} systems the owner-level amplification is
$+\Res{ai.ownerAmpMean}$ HHI (\Res{ai.ownerRatioMean}$\times$); at the outlet
level it is $+\Res{ai.outletAmpMean}$ (\Res{ai.outletRatioMean}$\times$).
Although the effect is smaller than at the outlet level, the result holds for \Res{ai.ownerCountries} of countries and under
every variant of the dataset we test (see the appendix). Hence, H2 is supported.

\paragraph{H3} \emph{AI concentration tracks ownership concentration}.
 Across countries, we correlate AI concentration and amplification with ownership HHI.
 Because countries with fewer owners score high on both by construction, we control for the number of owners and for traffic concentration.
Owner-level AI concentration correlates with
ownership HHI at $\rho = \Res{h44.ownerRawRho}$,
$\Res{h44.ownerPartialRho}$ after controlling for the number of owners in the
country ($p = \Res{h44.ownerPartialP}$), and $+0.35$ after controlling for both number of owners and traffic concentration.
This result holds under all \Res{h44.variantTotal} owner-set variants tested (see the appendix).
As such, among countries with comparable numbers of owners and levels of traffic concentration, the ones
whose ownership is more concentrated also see AI citations more concentrated by
owner.

While H3 is supported, analysis reveals an interesting relationship ($\rho=-0.32$*) between AI amplification and owner-level traffic HHI (Figure \ref{fig:ai_owner_amplification_b}).
Amplification ratios are higher where ownership is less concentrated, suggesting that AI systems may cause more media concentration to occur in markets with lower levels of media concentration.
As such, AI concentration is higher in absolute terms where ownership is concentrated, while the multiple by which it exceeds traffic concentration is larger where ownership is dispersed.

\paragraph{Ahrefs vs. Arena}
Table \ref{tab:ai-amplification} compares the amplification rates measured for a series of AI systems across the two datasets. We note that results on the arena dataset are underpowered, as coverage over our news domains is poor. However, on the $3,964$ overlapping domains, logged citation counts are well correlated ($\rho=0.536$). Similar to Ahrefs data, citation counts from Arena are significantly more concentrated at the outlet-level than traffic. We again see that amplification effects persist at the owner-level, but owner-level amplification is weaker compared to outlet-level across both datasets. While both datasets are subject to biases induced by data collection methodologies, corroboration between the two datasets reveals that the AI amplification of media concentration that we measure is not a product of these biases.

\subsection{State Media}

We demonstrate that state media ownership plays an important role in shaping journalism cultures by comparing to measures of press freedom. Then, we answer the question of whether AI citations show a preference toward state media.

\paragraph{Press freedom indices} We collect RSF indices on press freedom. \tabref{tab:state-sector-vs-freedom} shows that the proportion of news traffic that a country's state-sponsored media accounts for correlates negatively with RSF indices of press freedom \cite{rsf2026}. Counter to this, the proportion of traffic accounted for by public service broadcasting correlates positively with press freedom.

\input{tables/ai_state_h1}

\input{tables/state_sector_vs_freedom}

\paragraph{AI citation rates}
For each country, system, and class (state-controlled or public-service), we calculate an AI amplification rate (citations / traffic) and test the log ratio across countries with a Wilcoxon signed-rank test. On Ahrefs data, $p > 0.05$ for every system and class, and confidence intervals around the state-controlled ratios include 1 for every system (Table \ref{tab:ai-state-h1}).
At first glance, this finding is in contrast to recent research on the influence of state media on LLMs. It has been shown that, when a country has lower media freedom, LLMs tend to exhibit greater pro-government valence \emph{in that country's language} \citep{waight2026state}.
As such, we should not expect a preference to be visible in Ahrefs data, which is compiled using English prompts. The Arena dataset, however, is not subject to the same limitation.

\paragraph{H4} \emph{AI systems cite state media sources more than expected based on traffic levels when prompted in the relevant native language.}
In the Arena dataset, ~40\% of prompts are not in English. Restricting to users in a country, prompting in its national language, and citing that country's own outlets,
state-controlled media are cited \Res{arenaLang.ratio}$\times$ [\Res{arenaLang.ratioLo}, \Res{arenaLang.ratioHi}] more often than commercial outlets matched to the same traffic share: state-controlled outlets are cited at \Res{arenaLang.stateOE}$\times$ their traffic share, against \Res{arenaLang.placeboOE}$\times$ for the matched commercial bloc, across 2,961 citations in 16 countries.
Asked in a country's own language rather than in English, the same outlets' citations go to state-controlled media \Res{arenaLang.nativeShare} of the time against \Res{arenaLang.englishShare} (difference \Res{arenaLang.contrastDiff} [\Res{arenaLang.contrastLo}, \Res{arenaLang.contrastHi}]).
This is caused primarily by Russia and China.
There are 2.5$\times$ as many citations to Chinese state-controlled outlets in responses to Mandarin prompts as there are for English prompts. Likewise, Russian prompts provoke 5.3$\times$ more citations to Russian state media than English prompts do (Table \ref{tab:arena-state-language}).

\subsection{Media trust and AI amplification}

Here we ask whether the AI citations amplification effect is stronger in countries with lower media trust.
In countries where media trust is lower, if AI systems depend more heavily on smaller sets of outlets (for example, those that are well-known), a negative correlation between media trust and the AI amplification rates would follow.

For this, we use country level media trust scores from the Reuters Digital News Reports media trust survey instruments \citep{egan2026reuters}, and we find a significant negative correlation ($\rho = \Res{aiTrust.rho}$**, $n = \Res{aiTrust.n}$). Higher-trust countries see less owner-level AI amplification.
Furthermore, we find that media concentration does not explain the relationship between trust and AI amplification. While concentration correlates with amplification ($\rho = -0.429$**), media concentration and media trust are themselves unrelated ($\rho = \Res{aiTrust.concTrustRho}$), and so controlling for media concentration leaves the association between amplification and trust intact ($\rho = \Res{aiTrust.partialRho}$***).
So, the relationship between media trust and AI amplification persists independently of media concentration, suggesting that differences in media trust may be associated with how AI systems distribute citations across news outlets.
While this is one possible explanation, we refrain from making any claims that AI systems do in fact distribute citations according to media trust, and leave this for future work.

\section{Discussion}
\label{sec:discussion}

This work provides current and cross-national answers to questions around the number of organizations that operate in a country's media system and how much online traffic is concentrated among them. We motivate this work by showing how increased media concentration is associated with journalists' reports of organizational influence on editorial decisions (\autoref{fig:wjs_indicators}).
Our dataset provides a comparative baseline for the types of merger cases this paper opened with.
Take the Paramount--Warner Bros.\ Discovery litigation, for example.
Combining the two groups' outlets under one owner
raises our traffic-based estimate of US media concentration by \Res{policy.mergerDeltaPct}.
Now take the current market as the baseline, in which we assume that AI overviews mediate a negligible share of news traffic, and consider the case where all organic search traffic passes through Google AI Overviews.
In this scenario, with attention redistributed in proportion to Google AI Overview's citations, US media concentration rises by \Res{policy.aioUsOwnerDeltaPct}, roughly \Res{policy.aioVsMerger} times the Paramount merger's effect.
Furthermore, the effects are felt internationally, with the median country's owner-level HHI increasing by \Res{policy.aioOwnerDeltaPctMed} (in \Res{policy.aioOwnerHigherCount} of \Res{policy.aioOwnerCountries} countries).
Although we cannot yet conclude that traffic will come to follow citations, which requires longitudinal analysis, this scenario is indicative that platform and AI policy may now be more important than regulating mergers and acquisitions.

The strongest AI system amplifies concentration threefold more than the
weakest, and consulting several systems exposes a reader to meaningfully
more diversity than consulting one --- differences that dwarf most of the
cross-country variation in the market baseline. This inverts where policy
levers sit. The instruments of pluralism policy are national and aimed at
owners: licence caps, merger review, cross-ownership rules. The variation
that now matters most is between platforms, is set by citation behavior
rather than by ownership transactions, and is global in reach.
As such, FCC interventions in mergers and acquisitions between media conglomerates such as Paramount and Warner Bros. Discovery represent a national remedy applied to the ownership layer, while AI systems may have a more severe effect and are held accountable by no media regulator.
Exemplifying this issue is the fact that, according to our analysis, which assistant a reader uses may matter more for traffic concentration than which country they live in.

For the economics of digital news, AI systems impact media concentration through two mechanisms at once. The first is citation allocation, measured here: citations concentrate by owner beyond what attention concentration predicts.
The second is in traffic generation and click-through rates, where AI citations typically fail to deliver readers.
According to recent estimates, roughly one AI citation link in twenty is clicked, well below click-through in conventional search \citep{agarwal2026impact}.
As AI-mediated discovery displaces organic search, that ratio becomes a downwards pressure on referral traffic as a whole, and therefore on revenue.
In an industry defined by high fixed costs and low marginal costs \citep{noam2016media}, declining revenues will necessitate greater levels of efficiency on the production side.
Hence, the economies of scale that govern the organizational forms that media environments take on will strengthen, making media markets yet more concentrated in equilibrium.
Symptomatic of these efficiencies are the waves of persistent layoffs and reduced editorial output reported across the sector \citep{peterson_new_2023}.
The concern this work raises is that, through increasing traffic concentration (and decreasing traffic overall), AI systems as they exist today will indirectly exacerbate these issues.

We extend three lines of existing work. 
We reproduce Noam's four-firm ranking \citep{noam2016media}, while at the same time translating his media concentration program from national press and broadcast censuses to attention-weighted measurement of digital news. We show that, once ownership labels are attached, traffic concentration tracks ownership pressure reported by journalists. As such, monitors working from audience data alone (i.e., without ownership labels) may miss harms of media concentration. Finally, we extend the study of algorithmic curation and source diversity \citep{chakraborty_editorial_2018,abdollahpouri_toward_2021,choi_not_2022} from search and social feeds to generative systems, including their native-language preference for state-controlled media \citep{waight2026state}. Underlying all three is our ownership dataset. The lack of a consistent, cross-national news ownership dataset has been the chief obstacle to comparative analysis of media concentration, and we release the dataset \footnote{\url{https://doi.org/10.5281/zenodo.22782826}} and code so that concentration in new information systems can be monitored.

\section{Limitations}

The ownership labels are derived from Wikipedia
categories, Wikidata ownership fields and website links, and Wikipedia
infoboxes. As a result, underlying issues in these systems will carry forward into our dataset. For example, editor attention may vary across countries and language editions, and acquisitions may go unrecorded. Furthermore, our validation against the Medill dataset \citep{LocalNews2025} covers the United States only, where Wikipedia coverage is known to be stronger.
Where available, the integration of official data sources such as registries and regulator filings would extend that validation to countries where the data are weakest.
We note that, while our findings do reproduce on subsets of countries with better label coverage,
coverage is also negatively related to measured concentration ($\rho = \Res{coverage.vsHhiRho}$, $p = \Res{coverage.vsHhiP}$, $n = \Res{frame.ranked}$). Our labels cover \Res{coverage.domains} of domains and \Res{coverage.traffic} of traffic, so poorly covered countries may appear more concentrated than they are.

However, even given complete labels, the index uses traffic estimates and citations rather than editorial influence, advertising revenue or political power. This restricts
the production-economics argument in the Discussion. However, since it is known that informing voters is a critical
information need served by news media \citep{usher2023media}, election polling routinely asks which outlets shape voting
decisions \citep{fletcher2020mile}, and as a result, data to compute measures of editorial influence may be available. Reproducing our findings on influence based indices would strengthen the case for causal modeling of media concentration and its impacts.

These limitations have important implications for the AI citation analysis. Ahrefs data is from a
commercial panel whose prompt set is proprietary and English-
and United States-weighted. We therefore replicate on the AI Search Arena dataset
\citep{miroyan2026search}, whose prompts are organic user questions and
roughly 40\% of them not in English. Agreement across two panels makes an
artefact of either collection process less likely, though both remain
snapshots.
Moreover, AI systems cite few sources by
construction, so some elevation of concentration relative to traffic may be unavoidable, and our analysis does not account for the order or ranking of sources that get cited.
Finally, owner-level analysis of AI citations rests on a smaller set of \Res{ai.ownerCountries} countries,
because only \Res{ai.ownerDomains} of
\Res{ai.domains} AI-cited news domains carry an ownership label.
This issue is particularly acute for the Arena data, upon which any statistical claims are difficult to defend.
The AI findings are accordingly best read as generating hypotheses that
longitudinal data can test.

\section{Conclusion}

We release a cross-national news ownership dataset and use it to reveal that
media concentrations across \Res{frame.ranked} countries are higher than the corresponding outlet-level concentrations.
Our concentration indices correlate with the degree of commercial pressure reported by journalists, and after incorporating state media labels into the dataset, the indices reproduce external indicators of press freedom.
AI citation behavior then acts on that structure. AI citations are more concentrated than attention at the outlet level, and again once outlets are grouped by owner.
Asked in a country's own language, these systems cite state-controlled outlets \Res{arenaLang.ratio}$\times$ more than commercial outlets of the same size.
AI systems now allocate digital news attention at a scale comparable to the mergers regulators review, and we argue that traditional antitrust regulations should be extended to cover new AI systems. Without this, global trends of increasing media concentration are likely to accelerate.
The dataset, the labels and the full
analysis pipeline are released so that this work can be expanded upon.

\section{Acknowledgements}

We thank Dr. Zachary Metzger at the Medill Institute for providing ownership data for print outlets. This work was supported in part by the Office of Naval Research grant (N000142412568) and the Knight Foundation. 
The views and conclusions in this document are those of the authors and should not be interpreted as representing the official policies, either expressed or implied, of the Knight Foundation, Office of Naval Research, or the U.S. Government.

\bibliographystyle{ACM-Reference-Format}
\bibliography{aaai25}

\appendix
\input{sections/appendix_v2}

\end{document}

%% file: preamble.tex
\usepackage{ifthen}
\usepackage{booktabs}
\usepackage{subcaption}   
\usepackage{amsmath}
\usepackage{amssymb}

\newboolean{isanonymized}
\setboolean{isanonymized}{true} 

\newcommand{\anonymize}[2]{%
    \ifthenelse{\boolean{isanonymized}}{#1}{#2}%
}

\usepackage{resultsmacros}
\input{tex/results.tex}

\newcommand{\pBH}{\ensuremath{p_{\mathrm{BH}}}}

\newcommand{\figref}[1]{Figure~\ref{#1}}
\newcommand{\tabref}[1]{Table~\ref{#1}}

%% file: tex/results.tex
\DeclareRes{ai.ampPctMax}{86\%}
\DeclareRes{ai.ampPctMin}{67\%}
\DeclareRes{ai.domains}{4,584}
\DeclareRes{ai.models}{5}
\DeclareRes{ai.outletAmpMean}{445}
\DeclareRes{ai.outletRatioMean}{1.57}
\DeclareRes{ai.ownerAmpMean}{600}
\DeclareRes{ai.ownerCountries}{54}
\DeclareRes{ai.ownerDomains}{1,653}
\DeclareRes{ai.ownerRatioMax}{1.79}
\DeclareRes{ai.ownerRatioMean}{1.38}
\DeclareRes{ai.ownerRatioMin}{1.11}
\DeclareRes{ai.panelCountries}{199}
\DeclareRes{ai.rankCorr}{0.63}
\DeclareRes{ai.ratioMax}{2.08}
\DeclareRes{ai.ratioMin}{1.23}
\DeclareRes{ai.sameLeader}{21 of 76}
\DeclareRes{aiTrust.concTrustRho}{\ensuremath{+0.013}}
\DeclareRes{aiTrust.n}{40}
\DeclareRes{aiTrust.partialRho}{\ensuremath{-0.496}}
\DeclareRes{aiTrust.rho}{\ensuremath{-0.467}}
\DeclareRes{annotation.n}{200}
\DeclareRes{annotation.newsPrecision}{66.0\%}
\DeclareRes{arenaLang.contrastDiff}{10.7\%}
\DeclareRes{arenaLang.contrastHi}{12.9\%}
\DeclareRes{arenaLang.contrastLo}{8.4\%}
\DeclareRes{arenaLang.englishShare}{6.0\%}
\DeclareRes{arenaLang.nativeShare}{16.7\%}
\DeclareRes{arenaLang.placeboOE}{0.61}
\DeclareRes{arenaLang.ratio}{1.65}
\DeclareRes{arenaLang.ratioHi}{1.87}
\DeclareRes{arenaLang.ratioLo}{1.46}
\DeclareRes{arenaLang.stateOE}{1.01}
\DeclareRes{coverage.domains}{28.8\%}
\DeclareRes{coverage.traffic}{70.1\%}
\DeclareRes{coverage.vsHhiP}{0.002}
\DeclareRes{coverage.vsHhiRho}{\ensuremath{-0.347}}
\DeclareRes{cred.bestPartialPBH}{0.173}
\DeclareRes{cred.bestPartialRho}{\ensuremath{+0.243}}
\DeclareRes{cred.count.n}{77}
\DeclareRes{cred.count.pBH}{0.006}
\DeclareRes{cred.count.rho}{\ensuremath{+0.356}}
\DeclareRes{cred.lowOnly.p}{0.883}
\DeclareRes{cred.lowOnly.rho}{\ensuremath{+0.017}}
\DeclareRes{cred.nTests}{24}
\DeclareRes{cred.visits.pBH}{0.004}
\DeclareRes{cred.visits.rho}{\ensuremath{+0.386}}
\DeclareRes{desc.gbHhi}{1,877}
\DeclareRes{desc.gbOwners}{215}
\DeclareRes{desc.gbTopOne}{40\%}
\DeclareRes{desc.medianCrFour}{72\%}
\DeclareRes{desc.medianEffOwners}{5.8}
\DeclareRes{desc.medianHhi}{1,715}
\DeclareRes{desc.medianTopOne}{29\%}
\DeclareRes{desc.rankedCrFour}{80\%}
\DeclareRes{desc.rankedEffOwners}{4.5}
\DeclareRes{desc.reliableMinCoverage}{25\%}
\DeclareRes{desc.reliableMinDomains}{20}
\DeclareRes{desc.trHhi}{1,360}
\DeclareRes{desc.trOwners}{26}
\DeclareRes{filter.medianHhiAll}{2,115}
\DeclareRes{filter.medianHhiNews}{2,247}
\DeclareRes{filter.precision}{0.744}
\DeclareRes{filter.recall}{0.970}
\DeclareRes{filter.retained}{84.0\%}
\DeclareRes{filter.trafficPrecision}{0.831}
\DeclareRes{filter.unfilteredRho}{0.889}
\DeclareRes{frag.outletPressureP}{0.295}
\DeclareRes{frag.outletPressureRho}{\ensuremath{+0.144}}
\DeclareRes{frag.ownerPressureP}{0.007}
\DeclareRes{frag.ownerPressureRho}{\ensuremath{+0.359}}
\DeclareRes{frag.pressureN}{55}
\DeclareRes{frag.ratioMedian}{1.30}
\DeclareRes{frame.countries}{211}
\DeclareRes{frame.domains}{25,631}
\DeclareRes{frame.ranked}{77}
\DeclareRes{frame.reliable}{42}
\DeclareRes{frame.withOwner}{9,761}
\DeclareRes{frame.withTraffic}{22,928}
\DeclareRes{geo.n}{22,212}
\DeclareRes{geo.topFive}{93.2\%}
\DeclareRes{geo.topOne}{88.0\%}
\DeclareRes{h44.ownerPartialP}{\ensuremath{1.4 \times 10^{-4}}}
\DeclareRes{h44.ownerPartialRho}{\ensuremath{+0.496}}
\DeclareRes{h44.ownerRawRho}{\ensuremath{+0.766}}
\DeclareRes{h44.variantTotal}{5}
\DeclareRes{legacy.wikidata}{2,684}
\DeclareRes{lim.trustCountries}{48}
\DeclareRes{noam.cfour.n}{29}
\DeclareRes{noam.cfour.p}{0.011}
\DeclareRes{noam.cfour.rho}{\ensuremath{+0.465}}
\DeclareRes{noam.internalN}{20}
\DeclareRes{noam.internalRho}{\ensuremath{+0.149}}
\DeclareRes{noam.newspaper.n}{30}
\DeclareRes{noam.newspaper.p}{0.025}
\DeclareRes{noam.newspaper.pearson}{\ensuremath{+0.614}}
\DeclareRes{noam.newspaper.pearsonP}{\ensuremath{3.0 \times 10^{-4}}}
\DeclareRes{noam.newspaper.rho}{\ensuremath{+0.407}}
\DeclareRes{noam.online.n}{20}
\DeclareRes{noam.online.rho}{\ensuremath{+0.125}}
\DeclareRes{policy.aioOwnerCountries}{67}
\DeclareRes{policy.aioOwnerDeltaPctMed}{\ensuremath{+26.3\%}}
\DeclareRes{policy.aioOwnerHigherCount}{49}
\DeclareRes{policy.aioUsOwnerDeltaPct}{\ensuremath{+33.3\%}}
\DeclareRes{policy.aioVsMerger}{3.4}
\DeclareRes{policy.mergerDeltaPct}{\ensuremath{+9.9\%}}
\DeclareRes{spec.firstOwner}{0.955}
\DeclareRes{spec.minRho}{0.940}
\DeclareRes{spec.newsAllowlist}{0.940}
\DeclareRes{spec.rawOwners}{0.992}
\DeclareRes{spec.strictBlocklist}{0.973}
\DeclareRes{tail.medianOwnerPct}{80.1\%}
\DeclareRes{tail.medianTrafficPct}{47.8\%}
\DeclareRes{tail.oneOrTwoPct}{87.5\%}
\DeclareRes{tail.onePct}{76.9\%}
\DeclareRes{tail.oneTrafficPct}{27.5\%}
\DeclareRes{tail.owners}{3,540}
\DeclareRes{top.countries}{47}
\DeclareRes{top.crossBorder}{2}
\DeclareRes{top.public}{2}
\DeclareRes{trust.fullN}{47}
\DeclareRes{trust.fullP}{0.682}
\DeclareRes{trust.fullRho}{\ensuremath{-0.061}}
\DeclareRes{wjs.hhi.rho}{\ensuremath{+0.318}}
\DeclareRes{wjs.olsBeta}{\ensuremath{+0.302}}
\DeclareRes{wjs.olsN}{54}
\DeclareRes{wjs.olsP}{0.029}
\DeclareRes{wjs.olsRtwo}{0.478}
\DeclareRes{wjs.top1.rho}{\ensuremath{+0.359}}

%% file: tables/wjs_pressure_vs_concentration.tex
\begin{table}[t]
\centering
\small
\begin{tabular}{lrrr}
\toprule
Measure & Owners & Managers & Org \\
\midrule
Largest-owner share & +0.359** & +0.401** & +0.403** \\
HHI & +0.318* & +0.365** & +0.359** \\
Noam index & +0.242 & +0.290* & +0.257 \\
Top-4 owner share & +0.164 & +0.189 & +0.193 \\
\bottomrule
\end{tabular}
\caption{Spearman correlations across 55 countries between each measure of ownership concentration (rows) and each WJS rating of ownership pressure \citep{wjs3reports}: the influence journalists report from the owners of their news organisation, from its business managers, and the organizational-influence index ("Org"). Ratings run 1--5 with 5 most influential, so a positive value means journalists in more concentrated markets report more pressure. 
}
\label{tab:wjs-pressure-vs-concentration}
\end{table}

%% file: tables/ai_amplification.tex
\begin{table*}[t]
\centering
\small
\begin{tabular}{llrrrr}
\toprule
 & & \multicolumn{2}{c}{By outlet} & \multicolumn{2}{c}{By owner} \\
\cmidrule(lr){3-4} \cmidrule(lr){5-6}
Data & Model & AI $\div$ traffic HHI & $n_{\mathrm{AI}>\mathrm{traffic}}$ & AI $\div$ traffic HHI & $n_{\mathrm{AI}>\mathrm{traffic}}$ \\
\midrule
Ahrefs & Gemini & 2.08$\times$*** [1.65, 2.56] & 65 / 76 & 1.79$\times$*** [1.49, 2.00] & 47 / 53 \\
& Copilot & 1.60$\times$*** [1.38, 1.86] & 59 / 76 & 1.33$\times$*** [1.13, 1.57] & 42 / 54 \\
& Google & 1.57$\times$*** [1.36, 1.73] & 60 / 76 & 1.34$\times$** [1.14, 1.46] & 39 / 54 \\
& ChatGPT & 1.35$\times$*** [1.22, 1.63] & 61 / 76 & 1.33$\times$*** [1.16, 1.58] & 41 / 54 \\
& Grok & 1.23$\times$** [1.09, 1.38] & 51 / 76 & 1.11$\times$* [1.02, 1.31] & 35 / 54 \\
Arena & Perplexity & 1.76$\times$*** [1.47, 1.95] & 23 / 31 & 1.19$\times$* [0.99, 1.53] & 20 / 31 \\
& Gemini & 1.70$\times$ [1.02, 2.70] & 13 / 17 & 1.28$\times$ [0.99, 1.67] & 12 / 16 \\
& GPT 4.0 & 1.60$\times$* [1.23, 6.05] & 10 / 13 & 1.47$\times$* [1.09, 1.73] & 10 / 13 \\
\bottomrule
\end{tabular}
\caption{AI citation concentration against attention concentration, for each system, with outlets grouped by owner in the right-hand pair. Values are medians across countries; $n_{\mathrm{AI}>\mathrm{traffic}}$ counts the countries in which the system's citations are more concentrated than attention. Stars are a two-sided Wilcoxon signed-rank test of AI against traffic HHI across countries, uncorrected; Benjamini--Hochberg adjustment across the five systems leaves every star unchanged. `Google' is Google AI Overviews. `Sonar' is the Sonar-Reasoning-Pro perplexity model. Note that results on the Arena dataset \citep{miroyan2026search} are underpowered, with wide confidence intervals. * $p<0.05$, ** $p<0.01$, *** $p<0.001$.}
\label{tab:ai-amplification}
\end{table*}

%% file: tables/ai_state_h1.tex
\begin{table*}[t]
\centering
\small
\begin{tabular}{lllrrrrrr}
\toprule
Class & Data source & System & countries & cite : traffic & 95\% CI & amplified & suppressed & p \\
\midrule
public-service & Ahrefs (English panel) & ChatGPT & 15 & 0.51$\times$ & [0.26, 1.53] & 6 & 9 & .208 \\
 &  & Copilot & 15 & 0.22$\times$ & [0.12, 0.77] & 3 & 12 & \textbf{.008} \\
 &  & Gemini & 15 & 0.01$\times$ & [0.00, 1.63] & 5 & 10 & \textbf{.015} \\
 &  & Google & 15 & 0.57$\times$ & [0.01, 1.50] & 5 & 10 & .095 \\
 &  & Grok & 15 & 0.69$\times$ & [0.13, 1.36] & 7 & 8 & .277 \\
 & Arena (native language) & All systems & 3 & 0.82$\times$ & [0.01, 1.01] & 1 & 2 & -- \\
state-captured & Ahrefs (English panel) & ChatGPT & 26 & 0.92$\times$ & [0.36, 2.25] & 13 & 13 & .708 \\
 &  & Copilot & 26 & 0.50$\times$ & [0.25, 1.10] & 9 & 17 & \textbf{.027} \\
 &  & Gemini & 26 & 0.74$\times$ & [0.00, 1.70] & 9 & 17 & .053 \\
 &  & Google & 26 & 0.92$\times$ & [0.55, 2.12] & 12 & 14 & .727 \\
 &  & Grok & 26 & 0.99$\times$ & [0.54, 1.86] & 13 & 13 & .671 \\
 & Arena (native language) & All systems & 10 & \textbf{2.71$\times$} & [0.14, 9.87] & 7 & 3 & .492 \\
 &  & Perplexity & 9 & \textbf{2.67$\times$} & [0.01, 30.86] & 6 & 3 & .652 \\
  &  & Gemini & 5 & \textbf{1.61$\times$} & [0.01, 21.06] & 3 & 2 & 1.000 \\
 &  & GPT-4o & 4 & \textbf{1.15$\times$} & [0.01, 4.40] & 2 & 2 & 1.000 \\
\bottomrule
\end{tabular}
\caption{Is state media amplified or suppressed, per class, data source and AI system? For each country we take the class's share of a system's citations over its share of that country's news traffic; we report the median of that ratio across countries, how many countries fall on each side of parity, and a two-sided Wilcoxon signed-rank test of the log ratio against zero. Ahrefs rows come from answers to an English keyword panel; Arena rows are restricted to prompts in the country's own language, following previous work \citep{waight2026state}. Arena reaches too few countries to test public service broadcasting.}
\label{tab:ai-state-h1}
\end{table*}

%% file: tables/state_sector_vs_freedom.tex
\begin{table}[t]
\centering
\small
\begin{tabular}{lrrr}
\toprule
Class & countries & with sector & $\rho$ \\
\midrule
state-captured & 122 & 38 & $-$0.279** \\
independent public & 122 & 20 & +0.282**\\
pooled & 122 & 52 & $-$0.056  \\
\bottomrule
\end{tabular}
\caption{Correlates between each country's Reporters Without Borders (RSF) press freedom indicator and both the share of traffic received by that country's public service broadcasting and state-captured outlets.
}
\label{tab:state-sector-vs-freedom}
\end{table}

%% file: sections/appendix_v2.tex

\section{Validation, continued}

\input{tables/noam_correlations}

\paragraph{United Kingdom}
Using the same methodology, we gathered UK datasets and validated the coverage of UK news outlets on Wikidata with those that are listed in curated datasets maintained by the UK Public Interest News Foundation (PINF) \citep{public_interest_news_foundation}. Of the 1,438 Wikidata UK news domains we discover, 29\% are in the PINF dataset (or 482 news outlets).
Coverage varies by media type, with digital outlets best represented: our UK dataset covers 53.5\% of outlets which exist both in Print \& Online, 27.2\% of online-only outlets, 9.1\% of print-only titles, community radio 3.7\%, and local TV 2.9\%. In terms of ownership type, our dataset is biased toward corporate owned news outlets, covering 54\% of corporate titles, and only 9\% of independent outlets, which may partly reflect the large number of community radio stations PINF lists.

\paragraph{Country Assignment}
The country a domain is assigned to matches the top audience country
SimilarWeb reports for \Res{geo.topOne} of \Res{geo.n} domains, and is somewhere
in that domain's top five for \Res{geo.topFive}.
The dataset's country assignments are built from Wikipedia category trees, ccTLDs and
infobox locations. SimilarWeb's per-country audience shares are read by nothing
in that pipeline, so they are a genuinely held-out test rather than a consistency
check.
US traffic is inflated: a large share of measured web traffic originates in the United States whatever an outlet's market, so ``the top
audience country is the US'' is weak evidence of a US outlet. In cases where the US is marked as the top audience country by SimilarWeb, we count our country label as a match if it is either the US or the next highest country by traffic share.

\paragraph{Label Coverage and Inclusion Criteria}
Our inclusion criteria filter out a large number of domains pulled from Wikipedia. As determined via SimilarWeb traffic estimates, \Res{coverage.traffic} of the total traffic volume received by the news domains in our frame is accounted for by a domain that carries an ownership label.
Across the \Res{frame.ranked} ranked countries,
coverage and measured concentration correlate at $\rho =
\Res{coverage.vsHhiRho}$ ($p = \Res{coverage.vsHhiP}$), suggesting that inclusion criteria do impact the analysis. Thin coverage biases the index upward, because unlabeled traffic is dropped from the denominator.
However, changing the inclusion criteria and naming specifications does not move the concentration indices (see Specification Stability below).
As such, the analyses are not invalidated by reporting results on a subset of the traffic data observed (i.e. the subset for which we have ownership labels).

\section{Wikidata State-Sponsored Labels}
\label{app:wikidata-state}

The state-media classification used in the main text is the State Media
Monitor's \citep{dragomir2022state}. We also derive state labels from Wikidata,
which readers can reproduce from the released dataset.

\paragraph{Method}
A domain enters the pool on any evidence of public or state ownership: its own Wikidata entity, or an owner entity reached through the ownership properties (P127, P749, P123, P137), is typed as a public or state broadcaster; owner keys are propagated to every domain sharing the owner string; generic public or state naming patterns on the owner string (e.g., CBC, SRG SSR, TVNZ); and a small hand list. The pool is then partitioned by precedence into government non-media (keyed on \texttt{.gov}, \texttt{.mil} or \texttt{.edu} hosts and official gazettes), state-sponsored (positive evidence of state or ruling-party direction, optionally moderated by an RSF press-freedom gate), party press, and public-service as the remainder.

\paragraph{Analysis}
Table~\ref{tab:wikidata-state-breakdown} gives the class counts. The RSF gate
is the consequential choice, and the reason these labels are not
used for any result relating state media to press freedom: a label that reads
RSF cannot be used to test RSF. Without the gate, state-sponsored loses 70\% of its members and public-service grows by 46\%.

\begin{table}[t]
\centering
\small
\begin{tabular}{lrr}
\toprule
Class & shipped & rebuilt without RSF \\
\midrule
state-sponsored      & 285 & 85 \\
public-service       & 305 & 444 \\
party press          & 16  & 77 \\
government non-media & 54  & 54 \\
\bottomrule
\end{tabular}
\caption{A breakdown of Wikidata-derived labels for state media control.}
\label{tab:wikidata-state-breakdown}
\end{table}

Table~\ref{tab:wikidata-vs-smm} scores the labels against SMM's hand-compiled
census. They are precise and
incomplete: they rarely call an outlet state-controlled that SMM does not, but
they reach under half of what SMM lists. Dropping the RSF gate costs recall
without buying precision. This is why the main text reads SMM directly and these
labels are reported only here.

\begin{table}[t]
\centering
\small
\begin{tabular}{lrrr}
\toprule
Variant & precision & recall & F1 \\
\midrule
shipped (structural + RSF) & 0.969 & 0.461 & 0.625 \\
structural only            & 0.947 & 0.176 & 0.298 \\
rebuilt without RSF        & 0.862 & 0.123 & 0.215 \\
\bottomrule
\end{tabular}
\caption{Wikidata-derived state-sponsored labels against the SMM labels, on overlapping domains.}
\label{tab:wikidata-vs-smm}
\end{table}

\section{Language Subsets}
\label{app:language-subsets}

\input{tables/arena_state_language}

The Ahrefs citation counts are drawn per domain from a corpus of observed AI
answers, with no country or language parameter, and that corpus is built on a
predominantly English keyword base. The imbalance is visible in the frame:
english-facing outlets are 26.8\% of domains but take 61.9\% of citations,
against 40.1\% of traffic.
An outlet is english-facing if its Wikipedia infobox names English as a
publication language (available for 41\% of the frame), or its country's press
works substantially in English, or the domain is an explicit English edition
(\texttt{en.*}, \texttt{english.*}). The country rule over-includes, so the restriction is a
weak upper bound and the results below are correspondingly conservative.

\begin{table}[t]
\centering
\small
\begin{tabular}{llrrr}
\toprule
Level & Frame & $n$ & median amplification & $p$ \\
\midrule
outlet & full & 70 & $+537$ HHI (1.68$\times$) & $8.3\times10^{-9}$ \\
       & english-facing & 16 & $+424$ HHI (1.51$\times$) & $0.003$ \\
owner  & full & 70 & $+713$ HHI (1.38$\times$) & $1.4\times10^{-6}$ \\
       & english-facing & 16 & $+689$ HHI (1.38$\times$) & $0.016$ \\
\bottomrule
\end{tabular}
\caption{AI citation concentration against attention concentration, on the full
frame and on english-facing outlets only. Each figure is the mean of the five
per-model HHIs.}
\label{tab:english-facing-amplification}
\end{table}

Table~\ref{tab:english-facing-amplification} gives both frames. At the outlet
level the median country's AI ecosystem is $+424$ HHI points more
concentrated than its attention, against $+537$ on the full frame, and 14 of 16
countries amplify. At the owner level the restriction changes nothing at all:
1.38$\times$ on both frames, with 14 of 16 countries amplifying.

Restricting to native-language outlets does not recover a state-media
preference. Repeating the state-media test on
non-english-facing outlets only, state-controlled media are cited below their
traffic share by every system (median ratios 0.08--0.68$\times$ across 20
countries), two systems reach $p < .05$ uncorrected, and none survives
Benjamini--Hochberg correction across the ten tests (lowest $q = .094$).
What does recover a
preference is changing the language of the \emph{prompt} rather than the language
of the outlet (Table~\ref{tab:arena-state-language}), which is what H4 tests and
what \citet{waight2026state} report.

\section{Owner Layer versus Outlet Layer}
\label{app:layers}

On the same \Res{frag.pressureN} countries, the
ownership-pressure composite tracks owner-level concentration (largest-owner
share $\rho = \Res{frag.ownerPressureRho}$, $p = \Res{frag.ownerPressureP}$) and
does \emph{not} track outlet-level concentration, the HHI over news outlets
computed without using any ownership information at all
($\rho = \Res{frag.outletPressureRho}$, $p = \Res{frag.outletPressureP}$).
Table~\ref{tab:owner-variants-vs-pressure} gives the association under each way
of drawing the owner set.

\input{tables/owner_variants_vs_pressure}

The preceding checks establish that the index is measured properly. This
establishes that it measures something the outlet-level alternative does not,
against an external report of conditions inside newsrooms. Entered together against
media trust, neither layer carries an association, which is consistent with the
null reported in the credibility and trust section below.

\section{Specification Stability}
\label{app:dataset}

Every analytic specification agrees with the
primary country ranking at $\rho \geq \Res{spec.minRho}$: raw owner names
\Res{spec.rawOwners}, first-owner rather than fractional attribution
\Res{spec.firstOwner}, a stricter blocklist \Res{spec.strictBlocklist}, a
news-only allowlist \Res{spec.newsAllowlist}, and traffic floors higher still.
The single choice that moves the ranking is declining to filter non-news at all
(\Res{filter.unfilteredRho}), which is the point of the filter: it reorders
countries rather than rescaling them, and the median country HHI barely moves
(\Res{filter.medianHhiAll} to \Res{filter.medianHhiNews}).
Of the \Res{top.countries} countries where we identify a largest owner, only
\Res{top.crossBorder} corporate groups top more than one national market.
Per-country estimates are released with the dataset.

\section{External Validation}
\label{app:noam}

The comparison in the main text rests on \citet{noam2016media}'s four-firm
concentration ratio, the more robust of his two series and the one that survives
every specification we test. Two further results bound it. The like-for-like
online-news comparison is inconclusive ($\rho = \Res{noam.online.rho}$,
$n = \Res{noam.online.n}$), and his own two series agree with each other at only
$\rho = \Res{noam.internalRho}$ ($n = \Res{noam.internalN}$) on the countries
where both exist. Agreement with either is therefore evidence that our index
measures national ownership concentration; agreement with both would not be
available to any measure. On a log scale, our index and his newspaper-press HHI correlate at Pearson $r = \Res{noam.newspaper.pearson}$ ($p = \Res{noam.newspaper.pearsonP}$).
The full set of comparisons is in
Table~\ref{tab:noam-correlations}.

\section{AI Analysis}
\label{app:ai}

The panel is \Res{ai.domains} news domains across \Res{ai.panelCountries}
countries. A country enters the analysis with at least ten labeled domains,
rather than being cut to a fixed top-$N$: the owner-count gate already makes
markets comparable, and a fixed cut cost eight ranked countries while inflating
the amplification ratios it was meant to guard. Portals are excluded, as
elsewhere. AI concentration and traffic concentration are computed over the same
outlets, so a difference reflects how citations spread rather than which outlets
were sampled.

\paragraph{Sensitivity.} The amplification result holds under all three is-news
filter settings and at every top-$N$ cut from 10 to 30, though the effect shrinks
as $N$ grows --- adding smaller outlets adds units that AI rarely cites, which
raises the traffic baseline faster than the citation index. The association
between AI concentration and ownership concentration strengthens with $N$ rather
than weakening, so neither result is an artefact of the cut.

\paragraph{The AI hypothesis family.} H1--H3 are corrected
together within each level of analysis. Two tests are deliberately outside the family: the
traffic-versus-ownership correlation, which checks that the data behave as
expected rather than testing a hypothesis, and the trust tests, which address a
separate question.

\section{State and Public-Service Media}
\label{app:state}

Non-commercial ownership is split into four classes rather than treated as one,
because a public-service broadcaster, a state-sponsored outlet, a government
non-media site and a party organ are different objects and only the first is a
plurality institution in the regulatory sense.

On the Ahrefs panel, AI systems do not preferentially cite state-controlled or
public-service outlets, at either level of analysis and under the
citation-traffic elasticity control, and treating the state as a single owner
does not concentrate citations more than traffic. Those nulls are read as a size
effect rather than a state effect, because a size-matched commercial bloc is
\emph{more} suppressed than the state bloc: what looks like under-citation of
state media is under-citation of outlets of that size. They are also
underpowered, and the native-language result reported in the main text shows
they are a property of an English-prompted instrument.

\section{Credibility and Trust}
\label{app:negative}

Two further questions were asked of these data and are reported here because the
analyses exist and should not be repeated. Neither supports an argument in the
paper.

\paragraph{Credibility and consumption.} Concentration is positively associated
with the share of a country's news consumption going to outlets of low
\emph{or mixed} credibility ($\rho = \Res{cred.count.rho}$ by outlet count,
$\Res{cred.visits.rho}$ by visits; $\pBH = \Res{cred.count.pBH}$ and
$\Res{cred.visits.pBH}$, $n = \Res{cred.count.n}$), and that association is
robust to weighting and to frame. It does not survive a control for GDP per
capita: correcting for the \Res{cred.nTests} tests run in each frame,
\emph{none} of the GDP-controlled partials is significant, the strongest being
$\Res{cred.bestPartialRho}$ ($\pBH = \Res{cred.bestPartialPBH}$). Richer
countries are both less concentrated and consume less low-credibility news, and
that is enough to account for the raw association. There is no association at all
for clearly low-credibility outlets as distinct from low-or-mixed
($\rho = \Res{cred.lowOnly.rho}$, $p = \Res{cred.lowOnly.p}$).

\paragraph{Media trust.} We tested whether concentration relates to public trust
in news. It is not supported: there is no association across the full
\Res{trust.fullN}-country panel ($\rho = \Res{trust.fullRho}$,
$p = \Res{trust.fullP}$), and the positive association inside the canonical
Hallin and Mancini eighteen is specification-dependent and not significant on the
specification this paper adopts. It survives neither development controls, nor
four constructions of the public-ownership set, nor a between/within
decomposition by media system.

Both are limited by trust-survey coverage (\Res{lim.trustCountries} countries)
and by a single wave, and both need better-identified designs than we can bring
to them. Concentration in these data is consequential for the people producing
the news and, on this evidence, not measurably so for what audiences consume or
believe.

%% file: tables/noam_correlations.tex
\begin{table*}[t]
\centering
\small
\begin{tabular}{p{0.387\linewidth}rrrrr}
\toprule
Our measure vs Noam series & n & $\rho$ & p & $\rho$ SE (Fisher) & $\rho$ SE (bootstrap) \\
\midrule
HHI vs newspaper HHI & 30 & \textbf{+0.407} & \textbf{0.025} & 0.192 & 0.168 \\
HHI vs newspaper HHI (chapter alternative) & 30 & \textbf{+0.414} & \textbf{0.023} & 0.192 & 0.154 \\
HHI vs online-news HHI & 20 & +0.125 & 0.600 & 0.243 & 0.228 \\
Noam index vs Noam index & 24 & +0.404 & 0.050 & 0.218 & 0.193 \\
CR1 vs top-company share & 30 & +0.325 & 0.080 & 0.192 & 0.168 \\
CR4 vs newspaper C4 & 29 & \textbf{+0.465} & \textbf{0.011} & 0.196 & 0.171 \\
owners $\geq$1\% vs N of voices >1\% & 27 & \textbf{+0.408} & \textbf{0.035} & 0.204 & 0.172 \\
\bottomrule
\end{tabular}
\caption{Our index against each published Noam series.}
\label{tab:noam-correlations}
\end{table*}

%% file: tables/arena_state_language.tex
\begin{table*}[t]
\centering
\small
\begin{tabular}{llrrrrrr}
\toprule
Country & Language & RSF & state traffic share & citations & state & state O/E & commercial O/E \\
\midrule
RU & Russian & 25 & 0.287 & 1037 & 233 & 0.78 & 0.64 \\
CN & Chinese & 15 & 0.168 & 689 & 158 & 1.37 & 0.59 \\
VN & Vietnamese & 20 & 0.007 & 625 & 16 & 3.46 & 0.57 \\
BR & Portuguese & 64 & 0.000 & 258 & 0 & 0.00 & 0.62 \\
TR & Turkish & 29 & 0.013 & 74 & 2 & 2.13 & 0.90 \\
RO & Romanian & 66 & 0.002 & 69 & 4 & 25.64 & 0.38 \\
CZ & Czech & 84 & 0.035 & 55 & 11 & 5.69 & 0.17 \\
HU & Hungarian & 63 & 0.035 & 55 & 0 & 0.00 & 1.51 \\
UA & Ukrainian & 64 & 0.000 & 45 & 1 & 87.43 & 0.00 \\
IR & Persian & 16 & 0.009 & 28 & 1 & 3.80 & 1.14 \\
pooled & -- & -- & -- & 2961 & -- & 1.01 & 0.61 \\
\bottomrule
\end{tabular}
\caption{Do AI systems favour state media when asked in the country's own language? Each row is one country: citations to its own news outlets in answers to prompts in its national language, from the Search Arena dataset. `state O/E' is the observed number of citations to state-captured outlets over the number their share of the country's news traffic predicts; `commercial O/E' is the same for a bloc of commercial outlets matched to the state bloc's traffic share, which controls for AI under-citing large outlets in general. RSF is the Reporters Without Borders press-freedom score.}
\label{tab:arena-state-language}
\end{table*}

%% file: tables/owner_variants_vs_pressure.tex
\begin{table}[t]
\centering
\small
\begin{tabular}{p{0.478\linewidth}p{0.352\linewidth}r}
\toprule
owner set & largest-owner share vs pressure & n \\
\midrule
full dataset (primary) & \textbf{+0.359} (p = 0.007) & 55 \\
top-20 owners per country & \textbf{+0.350} (p = 0.009) & 55 \\
top-10 owners per country & \textbf{+0.340} (p = 0.011) & 55 \\
owner traffic $\geq$5M or $\geq$3 outlets & +0.293 (p = 0.087) & 35 \\
owner traffic $\geq$10M or $\geq$3 outlets & +0.220 (p = 0.261) & 28 \\
\emph{outlet-level fragmentation (any variant)} & \emph{+0.144 (p = 0.295)} & \emph{55} \\
\bottomrule
\end{tabular}
\caption{The pressure association under each way of drawing the owner set. Outlet-level fragmentation, which uses no owner label, is the contrast.}
\label{tab:owner-variants-vs-pressure}
\end{table}

%% file: aaai25.bib
@article{theine2025does,
  title={Does media ownership matter for journalistic content? A systematic scoping review of empirical studies},
  author={Theine, Hendrik and Bartsch, Julia and Troeger, Mandy},
  journal={International Journal of Communication},
  volume={19},
  pages={865--887},
  year={2025},
  publisher={USC ANNENBERG PRESS, UNIV SOUTHERN CALIFORNIA, KERCKHOFF HALL, 734 W ADAMS~…}
}

@article{similarweb,
  title={SimilarWeb},
  url={https://www.similarweb.com/corp/ourdata/},
  year={2026},
}

@book{hanitzsch2019worlds,
  title     = {Worlds of Journalism: Journalistic Cultures Around the Globe},
  editor    = {Hanitzsch, Thomas and Hanusch, Folker and Ramaprasad, Jyotika and de Beer, Arnold S.},
  publisher = {Columbia University Press},
  address   = {New York},
  year      = {2019}
}

@article{wjs3reports,
  title={Worlds of Journalism Study Wave 3 (2021-2025) Consolidated Dataset [Data set]},
  author={Lauerer, Corinna and Gr{\"u}newald, Mia and Hanitzsch, Thomas and Godole, Jonila and Zguri, Rrapo and Godole, Besmir and Amado, Adriana and Bongiovanni, Maximiliano and Rotelli, Nicol{\'a}s and Luque, Facundo and others},
  year={2025},
  doi= {10.5281/zenodo.15280738},
  publisher={CERN}
}

@book{noam2016media,
  title={Who owns the world's media?: media concentration and ownership around the world},
  author={Noam, Eli M},
  year={2016},
  publisher={Oxford University Press}
}

@online{ponsford2026journalists,
  author  = {Ponsford, Dominic},
  title   = {Which journalists and news outlets are most cited by AI answer engines?},
  year    = {2026},
  month   = apr,
  day     = {7},
  url     = {https://pressgazette.co.uk/news/journalists-cited-ai/},
  note    = {Press Gazette. Updated May 18, 2026}
}

@misc{bagdikian2004new,
  author = {Bagdikian, Ben H},
  title  = {The new media monopoly: A completely revised and updated edition with seven new chapters},
  year   = {2004},
  note   = {Beacon Press}
}

@article{hill2006measuring,
  title={Measuring media market diversity: Concentration, importance, and pluralism},
  author={Hill, Brian C},
  journal={Fed. Comm. LJ},
  volume={58},
  pages={169},
  year={2006},
  publisher={HeinOnline}
}

@inproceedings{abdollahpouri_toward_2021,
	year = {2021},
	location = {New York, {NY}, {USA}},
	title = {Toward the Next Generation of News Recommender Systems},
	isbn = {978-1-4503-8313-4},
	url = {https://dl.acm.org/doi/10.1145/3442442.3452327},
	doi = {10.1145/3442442.3452327},
	series = {{WWW} '21},
	pages = {402--406},
	booktitle = {Companion Proceedings of the Web Conference 2021},
	publisher = {Association for Computing Machinery},
	author = {Abdollahpouri, Himan and Malthouse, Edward C. and Konstan, Joseph A. and Mobasher, Bamshad and Gilbert, Jeremy},
	urldate = {2023-10-03},
	date = {2021-06-03},
}

@inproceedings{chakraborty_editorial_2018,
	year = {2018},
	location = {Republic and Canton of Geneva, {CHE}},
	title = {Editorial Algorithms: Optimizing Recency, Relevance and Diversity for Automated News Curation},
	isbn = {978-1-4503-5640-4},
	url = {https://dl.acm.org/doi/10.1145/3184558.3186937},
	doi = {10.1145/3184558.3186937},
	series = {{WWW} '18},
	shorttitle = {Editorial Algorithms},
	pages = {77--78},
	booktitle = {Companion Proceedings of the The Web Conference 2018},
	publisher = {International World Wide Web Conferences Steering Committee},
	author = {Chakraborty, Abhijnan and Luqman, Mohammad and Satapathy, Sidhartha and Ganguly, Niloy},
	urldate = {2023-10-03},
	date = {2018-04-23},
}

@inproceedings{choi_not_2022,
	location = {New York, {NY}, {USA}},
	title = {Do Not Read the Same News! Enhancing Diversity and Personalization of News Recommendation},
	isbn = {978-1-4503-9130-6},
	doi = {10.1145/3487553.3524936},
	series = {{WWW} '22},
	pages = {1211--1215},
	booktitle = {Companion Proceedings of the Web Conference 2022},
	publisher = {Association for Computing Machinery},
	author = {Choi, Seonghwan and Kim, Hyeondey and Gim, Manjun},
	urldate = {2023-10-03},
	date = {2022-08-16},
    year = {2022},
}

@article{ahrefs,
author={Ahrefs},
  title={SEO tools \& resources to grow your traffic},
  url={https://ahrefs.com/},
  year={2023},
}

@article{trappel_soaring_2022,
	year = {2022},
	title = {Soaring media ownership concentration: Comparing the effects of digitalisation on media pluralism and diversity},
	rights = {Creative Commons Attribution Non Commercial No Derivatives 4.0 International},
	doi = {10.48335/9789188855589-7},
	shorttitle = {Soaring media ownership concentration},
	author = {Trappel, Josef and Meier, Werner A.},
	urldate = {2023-10-11},
	date = {2022},
	langid = {english},
	note = {Publisher: Nordicom, University of Gothenburg},
}

@article{lidberg_distortion_2019,
  title={The distortion of the Australian public sphere: Media ownership concentration in Australia},
  author={Lidberg, Johan},
  journal={AQ-Australian Quarterly},
  volume={90},
  number={1},
  pages={12--20},
  year={2019},
  publisher={Australian Institute of Policy and Science Balmain, NSW}
}

@book{noam_media_2009,
  title={Media ownership and concentration in America},
  author={Noam, Eli},
  year={2009},
  publisher={Oxford University Press}
}

@incollection{sjovaag_media_2019,
	year = {2019},
	title = {Media Ownership and Journalism},
	isbn = {978-0-19-022861-3},
	booktitle = {Oxford Research Encyclopedia of Communication},
	author = {Sjøvaag, Helle and Ohlsson, Jonas},
	urldate = {2023-10-11},
	date = {2019-02-25},
	langid = {english},
	doi = {10.1093/acrefore/9780190228613.013.839},
}

@book{hallin2004comparing,
  title={Comparing media systems: Three models of media and politics},
  author={Hallin, Daniel C and Mancini, Paolo},
  year={2004},
  publisher={Cambridge university press}
}

@article{agarwal2026impact,
  title={The Impact of Google AI Overviews on Publisher Traffic and User Experience: Evidence from a Field Experiment},
  author={Agarwal, Saharsh and Sen, Ananya},
  year={2026}
}

@inproceedings{aggarwal2024geo,
  title={GEO: Generative engine optimization},
  author={Aggarwal, Pranjal and Murahari, Vishvak and Rajpurohit, Tanmay and Kalyan, Ashwin and Narasimhan, Karthik and Deshpande, Ameet},
  booktitle={Proceedings of the 30th ACM SIGKDD conference on knowledge discovery and data mining},
  pages={5--16},
  year={2024}
}

@article{tobitt2026news,
  author  = {Tobitt, Charlotte},
  title   = {Who’s suing AI and who’s signing: Encyclopedia Britannica and Merriam-Webster sue OpenAI following Perplexity lawsuit},
  journal = {Press Gazette},
  year    = {2026},
  month   = mar,
  day     = {24},
  url     = {https://pressgazette.co.uk/platforms/news-publisher-ai-deals-lawsuits-openai-google/},
  note    = {Updated April 7, 2026}
}

@article{chen2025generative,
  title={Generative engine optimization: How to dominate AI search},
  author={Chen, Mahe and Wang, Xiaoxuan and Chen, Kaiwen and Koudas, Nick},
  journal={arXiv preprint arXiv:2509.08919},
  year={2025}
}

@article{dragomir2022state,
  title={The state of state media},
  author={Dragomir, Marius and S{\"o}derstr{\"o}m, Astrid},
  journal={A Global Analysis of the Editorial Independence of State Media Based on the State Media Matrix},
  year={2022}
}

@article{egan2026reuters,
  title={Reuters Institute digital news report 2026},
  author={Egan, J and Robertson, CT and Ross Arguedas, A and Newman, N and Nielsen, RK and Mukherjee, M and Fletcher, R},
  doi={10.60625/risj-4vgf-s811},
  year={2026},
  publisher={Reuters Institute for the Study of Journalism}
}

@article{waight2026state,
  title={State media control influences large language models},
  author={Waight, Hannah and Yang, Eddie and Yuan, Yin and Messing, Solomon and Roberts, Margaret E and Stewart, Brandon M and Tucker, Joshua A},
  journal={Nature},
  pages={1--9},
  year={2026},
  publisher={Nature Publishing Group UK London}
}

@article{carragher2025misinformation,
  title={Misinformation resilient search rankings with webgraph-based interventions},
  author={Carragher, Peter and Williams, Evan M and Carley, Kathleen M},
  journal={ACM Transactions on Intelligent Systems and Technology},
  volume={16},
  number={1},
  pages={1--27},
  year={2025},
  publisher={ACM New York, NY}
}

@misc{rsf2026,
  author = {{Reporters sans frontières}},
  title = {World Press Freedom Index},
  year = {2026},
  url = {https://rsf.org/en/index}
}

@inproceedings{vrijenhoek2021recommenders,
  title={Recommenders with a Mission: Assessing Diversity in News Recommendations},
  author={Vrijenhoek, Sanne and Kaya, Mesut and Metoui, Nadia and M{\"o}ller, Judith and Odijk, Daan and Helberger, Natali},
  booktitle={Proceedings of the 2021 ACM SIGIR Conference on Human Information Interaction and Retrieval},
  pages={173--183},
  year={2021},
  publisher={ACM},
  doi={10.1145/3406522.3446019}
}

@inproceedings{vrijenhoek2022radio,
  title={{RADio}---Rank-Aware Divergence Metrics to Measure Normative Diversity in News Recommendations},
  author={Vrijenhoek, Sanne and B{\'e}n{\'e}dict, Gabriel and Gutierrez Granada, Mateo and Odijk, Daan and de Rijke, Maarten},
  booktitle={Proceedings of the 16th ACM Conference on Recommender Systems},
  pages={208--219},
  year={2022},
  publisher={ACM},
  doi={10.1145/3523227.3546780}
}

@article{scalvini2023making,
  title={Making Sense of Responsibility: A Semio-Ethic Perspective on TikTok's Algorithmic Pluralism},
  author={Scalvini, Marco},
  journal={Social Media + Society},
  volume={9},
  number={2},
  year={2023},
  doi={10.1177/20563051231180625}
}

@article{pierson2023governing,
  title={Governing Artificial Intelligence in the Media and Communications Sector},
  author={Pierson, Jo and Kerr, Aphra and Robinson, Stephen Cory and Fanni, Rosanna and Steinkogler, Valerie Eveline and Milan, Stefania and Zampedri, Giulia},
  journal={Internet Policy Review},
  volume={12},
  number={1},
  year={2023},
  doi={10.14763/2023.1.1683}
}

@misc{public_interest_news_foundation,
  title        = {Public Interest News Foundation},
  howpublished = {\url{https://www.publicinterestnews.org.uk/}},
  organization = {Public Interest News Foundation},
  address      = {4th Floor, 18 St. Cross Street, London, England, EC1N 8UN, UK},
  note         = {Registered Charity No. 1191397; Company No. 12320800},
  year         = {2019},
  url          = {https://www.publicinterestnews.org.uk/},
  urldate      = {2026-01-28}
}

@misc{LocalNews2024,
  author       = {Franklin, Tim},
  title        = {The State of Local News 2024: Expanding Deserts, Shifts in Ownership, and Expanded Digital Coverage},
  year         = {2024},
  howpublished = {Online report, Northwestern University},
  url          = {https://localnewsinitiative.northwestern.edu/projects/state-of-local-news/2024/},
  note         = {Accessed January 2026}
}

@misc{LocalNews2025,
  author       = {Franklin, Tim},
  title        = {The State of Local News 2025},
  year         = {2025},
  howpublished = {Online report, Northwestern University},
  url          = {https://localnewsinitiative.northwestern.edu/projects/state-of-local-news/2025/},
  note         = {Accessed January 2026}
}

@misc{GulbanCherryRoad2025,
  author       = {Caro, Mark},
  title        = {Paving the CherryRoad},
  howpublished = {\url{https://localnewsinitiative.northwestern.edu/posts/2025/05/12/jeremy-gulban-cherryroad-q-and-a/index.html}},
  year         = {2025},
  month        = may,
  note         = {Accessed: 2026-01-XX},
  institution  = {Local News Initiative, Northwestern University}
}

@article{peterson_new_2023,
	title = {The {New} {News} {Barons}: {Investment} {Ownership} {Reduces} {Newspaper} {Reporting} {Capacity}},
	volume = {707},
	issn = {0002-7162, 1552-3349},
	shorttitle = {The {New} {News} {Barons}},
	url = {http://journals.sagepub.com/doi/10.1177/00027162231211426},
	doi = {10.1177/00027162231211426},
	language = {en},
	number = {1},
	urldate = {2025-03-10},
	journal = {The ANNALS of the American Academy of Political and Social Science},
	author = {Peterson, Erik and Dunaway, Johanna},
	month = may,
	year = {2023},
	pages = {74--89},
}

@inproceedings{bisiani2025uktwitnewscor,
  title={Uktwitnewscor: A dataset of online local news articles for the study of local news provision},
  author={Bisiani, Simona and Gulyas, Agnes and Wihbey, John and Heravi, Bahareh},
  booktitle={Proceedings of the International AAAI Conference on Web and Social Media},
  volume={19},
  pages={2371--2384},
  year={2025}
}

@inproceedings{carragher2024detection,
  title={Detection and discovery of misinformation sources using attributed webgraphs},
  author={Carragher, Peter and Williams, Evan M and Carley, Kathleen M},
  booktitle={Proceedings of the International AAAI Conference on Web and Social Media},
  volume={18},
  pages={214--226},
  year={2024}
}

@inproceedings{miroyan2026search,
  title={Search arena: Analyzing search-augmented llms},
  author={Miroyan, Mihran and Wu, Tsung-Han and King, Logan and Li, Tianle and Pan, Jiayi and Hu, Xinyan and Chiang, Wei-Lin and Angelopoulos, Anastasios and Norouzi, Narges and Gonzalez, Joseph E and others},
  booktitle={International Conference on Learning Representations},
  volume={2026},
  pages={41109--41146},
  year={2026}
}

@article{yang2025news,
  title={News source citing patterns in ai search systems},
  author={Yang, Kai-Cheng},
  journal={arXiv preprint arXiv:2507.05301},
  year={2025}
}

@article{harcup2017news,
  title={What is news? News values revisited (again)},
  author={Harcup, Tony and O’neill, Deirdre},
  journal={Journalism studies},
  volume={18},
  number={12},
  pages={1470--1488},
  year={2017},
  publisher={Taylor \& Francis}
}

@article{usher2023media,
  title={Media policy for an informed citizenry: Revisiting the information needs of communities for democracy in crisis},
  author={Usher, Nikki and Darr, Joshua P and Napoli, Philip M and Miller, Michael L},
  journal={The ANNALS of the American Academy of Political and Social Science},
  volume={707},
  number={1},
  pages={8--20},
  year={2023},
  publisher={SAGE Publications Sage CA: Los Angeles, CA}
}

@techreport{fletcher2020mile,
  title       = {A Mile Wide, an Inch Deep: Online News and Media Use in the 2019 UK General Election},
  author      = {Fletcher, Richard and Newman, Nic and Schulz, Anne},
  institution = {Reuters Institute for the Study of Journalism},
  year        = {2020},
  doi         = {10.60625/risj-205t-c448}
}

@inproceedings{aljebreen2024analysis,
  author={Aljebreen, Abdullah and Meng, Weiyi and Dragut, Eduard C.},
  title={Analysis and Detection of ``Pink Slime'' Websites in Social Media Posts},
  booktitle={Proceedings of the ACM Web Conference 2024 (WWW '24)},
  year={2024}, pages={2572--2581}, doi={10.1145/3589334.3645588}, publisher={ACM}}

@inproceedings{papadogiannakis2023who,
  author={Papadogiannakis, Emmanouil and Papadopoulos, Panagiotis and Markatos, Evangelos P. and Kourtellis, Nicolas},
  title={Who Funds Misinformation? A Systematic Analysis of the Ad-related Profit Routines of Fake News Sites},
  booktitle={Proceedings of the ACM Web Conference 2023 (WWW '23)},
  year={2023}, pages={2765--2776}, doi={10.1145/3543507.3583443}, publisher={ACM}}

@inproceedings{chen2024global,
  author={Chen, Xi and Hale, Scott A. and Jurgens, David and Samory, Mattia and Zuckerman, Ethan and Grabowicz, Przemyslaw A.},
  title={Global News Synchrony and Diversity During the Start of the {COVID}-19 Pandemic},
  booktitle={Proceedings of the ACM Web Conference 2024 (WWW '24)},
  year={2024}, pages={2639--2650}, doi={10.1145/3589334.3645645}, publisher={ACM}}

@inproceedings{robertson2018auditing,
  author={Robertson, Ronald E. and Lazer, David and Wilson, Christo},
  title={Auditing the Personalization and Composition of Politically-Related Search Engine Results Pages},
  booktitle={Proceedings of the 2018 World Wide Web Conference (WWW '18)},
  year={2018}, pages={955--965}, doi={10.1145/3178876.3186143}, publisher={ACM}}

@inproceedings{le2019measuring,
  author={Le, Huyen and Maragh, Raven and Ekdale, Brian and High, Andrew and Havens, Timothy and Shafiq, Zubair},
  title={Measuring Political Personalization of {Google} News Search},
  booktitle={The World Wide Web Conference (WWW '19)},
  year={2019}, pages={2957--2963}, doi={10.1145/3308558.3313682}, publisher={ACM}}

@inproceedings{urman2021auditing,
  author={Urman, Aleksandra and Makhortykh, Mykola and Ulloa, Roberto},
  title={Auditing Source Diversity Bias in Video Search Results Using Virtual Agents},
  booktitle={Companion Proceedings of the Web Conference 2021 (WWW '21 Companion)},
  year={2021}, pages={232--236}, doi={10.1145/3442442.3452306}, publisher={ACM}}

@inproceedings{sayyidali2026are,
  author={Sayyid-Ali, Abdur-Rahman Ibrahim and Khan, Daanish Uddin and Bhatti, Naveed Anwar},
  title={Are {LLM} Web Search Engines Sustainable? A Web-Measurement Study of Real-Time Fetching},
  booktitle={Proceedings of the ACM Web Conference 2026 (WWW '26)},
  year={2026}, pages={2083--2093}, doi={10.1145/3774904.3792278}, publisher={ACM}}

@inproceedings{chen2026unveiling,
  author={Chen, Pei and Hong, Geng and Wu, Xinyi and Wu, Mengying and Zhu, Zixuan and Liu, Mingxuan and Liu, Baojun and Zhang, Mi and Yang, Min},
  title={Unveiling the Resilience of {LLM}-Enhanced Search Engines against Black-Hat {SEO} Manipulation},
  booktitle={Proceedings of the ACM Web Conference 2026 (WWW '26)},
  year={2026}, pages={2962--2973}, doi={10.1145/3774904.3792399}, publisher={ACM}}

@inproceedings{chouaki2024what,
  author={Chouaki, Salim and Chakraborty, Abhijnan and Goga, Oana and Zannettou, Savvas},
  title={What News Do People Get on Social Media? Analyzing Exposure and Consumption of News through Data Donations},
  booktitle={Proceedings of the ACM Web Conference 2024 (WWW '24)},
  year={2024}, pages={2371--2382}, doi={10.1145/3589334.3645399}, publisher={ACM}}

@inproceedings{spangher2020characterizing,
  author={Spangher, Alexander and Ranade, Gireeja and Nushi, Besmira and Fourney, Adam and Horvitz, Eric},
  title={Characterizing Search-Engine Traffic to {Internet Research Agency} Web Properties},
  booktitle={Proceedings of The Web Conference 2020 (WWW '20)},
  year={2020}, pages={2253--2263}, doi={10.1145/3366423.3380290}, publisher={ACM}}
